\documentclass[fleqn]{fac}

\journal{Preprint for arXiv}

\usepackage{times}
\usepackage{color}
\usepackage{stmaryrd}
\usepackage{amssymb}
\usepackage{amsmath}
\usepackage{ifthen}
\usepackage[pdftex]{graphicx}
\usepackage{imakeidx}
\makeindex
\usepackage{url}
\usepackage{calc}

\newtheorem{definition}{Definition}

\usepackage{laws}

\newcommand{\draftonly}[1]{}

\usepackage[colorinlistoftodos]{todonotes}

\newenvironment{relational}{\it\pagebreak[1]\par\noindent\rule{\columnwidth}{0.5pt}\par\noindent}{\nopagebreak\rule{\columnwidth}{0.5pt}}

\newcommand{\figurerule}{\rule{\textwidth}{0.5pt}}

\renewenvironment{proof}{\noindent \textit{Proof.}}{\noindent$\Box$\par}

\newcommand{\labelaxiom}[1]{\label{axiom-#1}\index{Axiom!#1|AxiomDef{\ref{axiom-#1}}}}
\makeatletter
\def\refaxiom{\@ifnextchar*{\@refaxiom}{\@@refaxiom}}
\def\@refaxiom*#1{\ref{axiom-#1}\index{Axiom!#1|LawUse}}
\def\@@refaxiom#1{axiom~(\ref{axiom-#1})\index{Axiom!#1|LawUse}}
\makeatother

\newcommand{\labelproperty}[1]{\label{property-#1}\index{Property!#1|PropertyDef{\ref{property-#1}}}}
\makeatletter
\def\refproperty{\@ifnextchar*{\@refproperty}{\@@refproperty}}
\def\@refproperty*#1{\ref{property-#1}\index{Property!#1|LawUse}}
\def\@@refproperty#1{property~(\ref{property-#1})\index{Property!#1|LawUse}}
\makeatother

\newcommand{\labeldefinition}[1]{\label{def-#1}\index{Definition!#1|DefinitionDef{\ref{def-#1}}}}
\makeatletter
\def\refdefinition{\@ifnextchar*{\@refdefinition}{\@@refdefinition}}
\def\@refdefinition*#1{\ref{def-#1}\index{Definition!#1|LawUse}}
\def\@@refdefinition#1{definition~(\ref{def-#1})\index{Definition!#1|LawUse}}
\makeatother

\newcommand{\Pre}[1]{\{#1\}}
\newcommand{\Rely}{\mathop{\Keyword{rely}}}
\makeatletter
\def\rely{\@ifnextchar*{\@rely}{\@@rely}}
\def\@rely*#1#2#3{\Rely #2 \spot #3_{#1}}
\def\@@rely#1#2{\Rely #1 \spot #2}

\def\RelyTerm{\@ifnextchar*{\@RelyTerm}{\@@RelyTerm}}
\def\@RelyTerm*#1#2#3{#2 \entails \Stopped(#3,r_x \lor #1))}
\def\@@RelyTerm#1#2#3{#2 \equiv \Stopped(#3,r_x \lor #1))}
\makeatother
\newcommand{\Seq}{\mathbin{;}}
\newcommand{\SSeq}{\,}
\newcommand{\cat}{\mathbin{\raise 0.8ex\hbox{$\frown$}}}

\newcommand{\atomicrel}[1]{\langle#1\rangle}
\newcommand{\FinSkipIter}{^{\varoast}}
\newcommand{\FinGuar}[1]{\atomicrel{#1}\FinSkipIter}
\newcommand{\omegaskip}{\circledcirc}
\newcommand{\InfSkipIter}{^{\omegaskip}}
\newcommand{\InfGuar}[1]{\atomicrel{#1}\InfSkipIter}
\newcommand{\Nil}{\Keyword{nil}}

\newcommand{\id}{\mathsf{id}}
\newcommand{\sdefs}{\mathrel{\widehat=}}
\newcommand{\spot}{\mathrel{{\cdot}}}
\newcommand{\where}{\mathrel{|}}
\renewcommand{\implies}{\mathrel{\Rightarrow}}

\newcommand{\universalrel}{\mathsf{univ}}

\newcommand{\relint}{\cap}
\newcommand{\relcontained}{\subseteq}
\renewcommand{\iff}{\mathrel{\Leftrightarrow}}
\newenvironment{refine}{\begin{displaymath}\begin{array}{l}}{\end{array}\end{displaymath}}
\newcommand{\Altx}{,~~}

\newcommand{\Command}{\mathit{Com}}
\newcommand{\refines}{\mathrel{\sqsupseteq}}
\newcommand{\refsto}{\mathrel{\sqsubseteq}}
\newcommand{\nondet}{\mathbin{\sqcap}}
\newcommand{\Nondet}{\mathop{\textstyle\bigsqcap}}
\newcommand{\angelic}{\mathbin{\sqcup}}
\newcommand{\Angelic}{\mathop{\textstyle\bigsqcup}}
\newcommand{\finiterec}{\nu}
\newcommand{\infiniterec}{\mu}

\newcommand{\ifinv}[2]{#2}

\newcommand{\strictconjunction}{weak conjunction}
\newcommand{\Strictconjunction}{Weak conjunction}
\newcommand{\together}{\mathbin{\doublecap}}
\newcommand{\quotient}{\mathbin{/\!\!/}}

\newcommand{\Env}{\Keyword{env}}
\newcommand{\envc}[1]{(\Env~#1)}

\newcommand{\PrefixClose}[1]{\mathit{prefixes}(#1)}
\newcommand{\Trace}{\mathit{Trace}}

\newcommand{\Keyword}[1]{\mathsf{\mathbf{#1}}}
\newcommand{\Magic}{\Keyword{\top}}
\newcommand{\Abort}{\Keyword{\bot}}
\newcommand{\botstate}{\bot}
\newcommand{\Chaos}{\Keyword{chaos}}

\newcommand{\entails}{\Rrightarrow}

\newcommand{\FinIter}{^{\star}}
\newcommand{\itkleene}[1]{#1^{\star}}

\newcommand{\FinOrInfIter}{^{\circ}}
\newcommand{\itomega}[1]{#1\FinOrInfIter}

\newcommand{\pstepl}[1]{\Pi(#1)}
\newcommand{\estepl}[1]{{\cal E}(#1)}
\newcommand{\termd}{\checkmark}

\newcommand{\cpstepd}{\boldsymbol{\pi}}
\newcommand{\cestepd}{\boldsymbol{\epsilon}}
\newcommand{\cestepbotd}{\cestepd_{\botstate}}

\newcommand{\cpstep}[1]{\cpstepd(#1)}
\newcommand{\cestep}[1]{\cestepd(#1)}
\newcommand{\cestepbot}[1]{\cestepbotd(#1)}
\newcommand{\cgdd}{\boldsymbol{\tau}}
\newcommand{\cgd}[1]{\cgdd(#1)}

\newcommand{\While}{\mathop{\Keyword{while}}}
\newcommand{\Do}{\mathop{\Keyword{do}}}

\newcommand{\Skip}{\Keyword{skip}}
\newcommand{\Term}{\Keyword{term}}

\makeatletter
\def\Spec{\@ifnextchar*{\@Spec}{\@@Spec}}
\def\@Spec*#1#2#3{\ifx\@empty#1\else#1\colon\fi
   [#2\ifx\@empty#2\else,~\fi#3]}
\def\@@Spec#1#2#3{\ifx\@empty#1\else
   \begin{array}{@{}l@{}}#1\colon\end{array}\!\!\fi%
   \left[\begin{array}{@{}c@{}}#2\end{array}\ifx\@empty#2\else,~\fi
   \begin{array}{@{}c@{}}#3\end{array}\right]}

\newcommand{\ChainRel}[1]{\crcr \noalign{\penalty\interdisplaylinepenalty}
  \hspace*{-1em}#1
  \@ifnextchar*{\@ChainRelCommment}{}}
\newcommand{\ChainRelFormat}[1]{\mbox{~~~#1}}
\def\@ChainRelCommment*[#1]{\ChainRelFormat{#1}
  \crcr \noalign{\penalty\interdisplaylinepenalty}}
\newcommand{\StartRef}[1]{\hspace*{-1.5em}(\ref{#1}) \refsto
  \@ifnextchar[{\@StartRefCommment}{}}
\def\@StartRefCommment[#1]{\mbox{#1}
  \crcr \noalign{\penalty\interdisplaylinepenalty}}
\makeatother
\newcommand{\Implies}{\ChainRel{\implies}}
\newcommand{\IFF}{\ChainRel{\iff}}

\newcommand{\ImpliedBy}{\ChainRel{\Leftarrow}}
\newcommand{\Refsto}{\ChainRel{\refsto}}

\newcommand{\Equals}{\ChainRel{=}}

\edef\today{\number\day\ \ifcase\month\or
  January\or February\or March\or April\or May\or June\or
  July\or August\or September\or October\or November\or December\fi
  \ \number\year}
\newcounter{Hours}
\newcounter{Minutes}
\newcommand{\CurrentTime}{%
 \ifthenelse{\value{Hours}<10}{0}{}\theHours:%
 \ifthenelse{\value{Minutes}<10}{0}{}\theMinutes}
\newcommand{\runningdate}{\draftonly{(\today\ DRAFT)}}

\begin{document}

\newcommand{\Reviewers}[1]{}

\title[Generalised rely-guarantee concurrency \runningdate]{Generalised rely-guarantee concurrency: An algebraic foundation}
\author[Ian J. Hayes \runningdate]{Ian J. Hayes \\
School of Information Technology and Electrical Engineering, The University of Queensland, Australia
}
\correspond{Ian.Hayes@itee.uq.edu.au}
\makecorrespond

\maketitle

\begin{abstract}
The rely-guarantee technique allows one to reason compositionally about concurrent programs.
To handle interference the technique makes use of rely and guarantee conditions,
both of which are binary relations on states.
A rely condition is an assumption that the environment performs
only atomic steps satisfying the rely relation
and a guarantee is a commitment that every atomic step 
the program makes satisfies the guarantee relation.
In order to investigate rely-guarantee reasoning more generally,
in this paper we allow interference to be represented by a 
process rather than a relation
and hence derive more general rely-guarantee laws.
The paper makes use of a \strictconjunction\ operator between processes,
which generalises a guarantee relation to a guarantee process,
and introduces a rely quotient operator, 
which generalises a rely relation to a process.
The paper focuses on the algebraic properties of the general 
rely-guarantee theory.
The Jones-style rely-guarantee theory can be interpreted  
as a model of the general algebraic theory
and hence the general laws presented here hold for that theory. 
\end{abstract}


\section{Introduction}

\paragraph{Rely and guarantee conditions.}

The rely-guarantee technique of Jones \cite{Jones81d,jon83a,jones96a}
provides a compositional approach to reasoning about concurrent programs.
With hindsight, it is obvious that to achieve compositional handling of concurrency, 
it is necessary to have some way of recording information about interference.
This paper generalises the way that interference is recorded.
To allow reasoning about a process $c$ in isolation,
Jones used a \emph{rely} condition $r$,
that is a binary relation on states.
Every atomic step of the environment of $c$ is assumed to satisfy the rely condition $r$
between its before and after states.
Any process running in parallel with $c$ also has a rely condition 
and hence process $c$ will need to ensure every atomic program step it makes satisfies 
the rely conditions of all processes in its environment.
To represent this Jones uses a \emph{guarantee} condition $g$,
that is also a binary relation on states.
Every atomic step of $c$ must satisfy $g$
and the relation $g$ should be contained in the rely condition of every process
in the environment of $c$.
Jones records a rely-guarantee specification by generalising the judgements of 
Hoare logic \cite{Hoare69a} to a quintuple of the form,
\begin{eqnarray}\labelproperty{quintuple}
  \Pre{p,r} ~c~ \Pre{g,q}~.
\end{eqnarray}
The process $c$ satisfies the quintuple if, 
under the assumption that
the initial state satisfies $p$ and 
every atomic step made by the environment satisfies $r$ between its before and after states,
every possible execution of $c$ 
ensures that every atomic program step made by $c$ satisfies $g$, 
and the initial and final states of the overall execution of $c$
satisfy the relational postcondition $q$.

\paragraph{Refinement calculus.}

This paper uses a refinement calculus approach \cite{Bac81a,BackWright98,TSS,Morgan94,ATBfSRatPC}
rather than Hoare logic because it allows for simpler presentation of algebraic laws of programming \cite{HoareHayesEtcFull87}.
Refinement of one command $c$ by another $d$ is written ``$c \refsto d$'' and 
is read ``$c$ is refined (implemented) by $d$''.
The refinement calculus introduces 
a postcondition specification command $\Spec{}{}{q}$ in which the postcondition $q$ is a binary relation on states,
and a precondition command $\Pre{p}$ in which the precondition $p$ is a set of states.
The refinement $\Pre{p} \Seq \Spec{}{}{q} \refsto d$ means $d$ achieves the postcondition $q$ between
its before-state and after-state, provided its before-state satisfies $p$.
As an abbreviation the sequential composition operator ``$\Seq$'' may be elided 
so that the above may be written $\Pre{p} \SSeq \Spec{}{}{q}$.

\paragraph{Generalised rely-guarantee.}

The main contribution of this paper is to generalise a rely condition $r$ to 
a process $i$ specifying the assumed behaviour of interference from the environment.
The actual environment should satisfy (i.e.\ refine) the process specification $i$.
Similarly, the guarantee condition $g$ is generalised to a process $j$
to be ``guaranteed'' by the implementation.
The process that behaves as a process $c$ as well as respecting the guarantee process $j$
is represented by their \strictconjunction\ $j \together c$,
which is the process that behaves as both $j$ and $c$ unless one of them aborts.%
\footnote{Earlier publications referred to \strictconjunction\ as \emph{strict} conjunction
but the new name is preferred because the operator is weaker than the (strong) conjunction operator 
that requires both its operands to abort for it to abort.}
A Jones-style guarantee condition $g$ on a terminating command $c$ is represented by 
the process $\FinGuar{g} \together c$,
where 
$\atomicrel{g}$ represents a command that can perform a single atomic program step 
for which the before and after states satisfy $g$
and
$\FinGuar{g}$ is the process 
that iterates the atomic step $\atomicrel{g}$ any finite number
of times, zero or more.
An example of a guarantee process that cannot be expressed as a guarantee condition is
the sequential composition 
$\FinGuar{\id} \SSeq \atomicrel{g} \SSeq \FinGuar{\id}$,
in which $\id$ is the identity relation.
It guarantees that a step satisfying $g$ occurs exactly once but allows stuttering 
steps before and after.
The closest guarantee condition is $g \cup \id$ but that allows any number, zero of more, of steps satisfying $g \cup \id$.
Section~\ref{section:conjunction} explores the \strictconjunction\ operator and 
its relationship to Jones-style guarantee conditions~\cite{FACJexSEFM-14}.

\paragraph{Rely quotients.}

To specify a process that refines (implements) $c$,
while relying on its environment refining process $i$,
a rely quotient operator $c \quotient i$ is introduced.
The rely quotient $c \quotient i$ when run in parallel with $i$ implements $c$,
\begin{eqnarray*}
  c  & \refsto & (c \quotient i) \parallel i~.
\end{eqnarray*}
The operator ``$\quotient$'' is chosen to be similar in appearance to the division operator, 
where in this context ``$\parallel$'' takes on a role similar to multiplication.
\label{division}Taking ``$x \quotient y$'' as the ceiling of their integer division $\lceil x/y \rceil$ gives the best analogy:
$x \leq \lceil x/y \rceil \times y$.
A terminating process specification $c$ with a Jones-style rely condition $r$ 
is represented by the quotient $c \quotient \FinGuar{r}$,
where $\FinGuar{r}$ represents the environment process,
all atomic steps of which satisfy $r$.
Section~\ref{section:rely} explores the properties of the rely quotient operator.
Given the \strictconjunction\ and rely quotient operators, the Jones quintuple (\refproperty*{quintuple})
is equivalent to the following refinement.
\begin{eqnarray}\labelproperty{quintuple-as-refinement}
  \Pre{p} \SSeq (\FinGuar{g} \together (\Spec{}{}{q} \quotient \FinGuar{r})) & ~\refsto~ & c
\end{eqnarray}

\paragraph{Concurrency.}

The parallel introduction law of Jones makes use of both rely and guarantee conditions.
In the more general theory presented here, 
\strictconjunction\ takes on the role of a guarantee 
and the rely quotient takes on the role of a rely condition.
Both generalised operators are used 
to give a general version a law for introducing a parallel composition,
which has a surprisingly simple and elegant proof (see Section~\ref{section:parallel}).

\paragraph{Distribution laws.}

Section~\ref{section:rely-distribution} examines the distribution properties of 
the rely quotient operator over the other operators.
In some cases the general distribution laws for \strictconjunction\ and rely quotient
require provisos.
However, in the relational rely-guarantee model
the provisos are all valid and hence the distribution properties hold without proviso.
In the general theory the provisos are explicit and 
hence it is possible to explore alternatives to Jones-style rely-guarantee
that allow more expressive rely conditions.

\paragraph{Relationship to relational rely-guarantee.}

Exploring the theory more generally 
leads to simpler laws 
that can be specialised to the relational model.
As an example consider the nesting of two rely processes $i$ and $j$, i.e.\ $(c \quotient j) \quotient i$. 
That corresponds to handling concurrent interference from both $i$ and $j$
and is equivalent to $c \quotient (i \parallel j)$,
i.e.\ an effective rely process of $i \parallel j$.
A relational rely condition of $r$ corresponds to a rely process of $\FinGuar{r}$
and the nesting of two such processes for rely conditions of $r_0$ and $r_1$
corresponds to the rely process of 
$\FinGuar{r_0} \parallel \FinGuar{r_1}$,
however, 
this process is equivalent to $\FinGuar{r_0 \lor r_1}$,
corresponding to a relational rely of $r_0 \lor r_1$.
This shows how the well known relational rely-guarantee rule, 
that the effective rely of nested relational rely conditions is their disjunction,
can be derived from the more general view that the effective rely process 
of nested rely processes is their parallel composition.

Section~\ref{section:Jones-rely} explores the relationship 
of the more general theory to the Jones-style relational guarantee and rely conditions.
The relational rely-guarantee theory of Jones \cite{jones96a}
is a model of the general algebraic theory presented in this paper
and hence the laws developed in the general theory are also valid for Jones' theory.

Section~\ref{section:fair-parallel} examines fair parallel and its impact on the rely quotient operator.

\paragraph{Contributions.}

The main contribution of this paper is to generalise rely and guarantee conditions from relations to
arbitrary processes.
In order to make our results as widely applicable as possible,
we have based our theory on a relatively small set of definitions and axioms.
Any model, such as the relational rely-guarantee model, that satisfies the axioms
can then make use of all the laws proved here.

Our core theory adds two specification operators, \strictconjunction\ and rely quotient,
to the operators of a simple parallel programming language.
The \strictconjunction\ operator allows guarantees to be imposed on a process \cite{HayesJonesColvin14TR}.
The rely quotient operator introduced in this paper allows rely conditions to be generalised to processes.
There are a number of advantages of exploring the more general operators.
Both \strictconjunction\ and rely quotient have simple algebraic properties
and this leads to simple and elegant proofs of laws involving these operators.
The approach leads to a nice separation of concerns 
because properties of \strictconjunction\ (guarantees) and rely quotient can
be developed separately
and then combined to give 
generalised equivalents of the main laws used for standard rely-guarantee refinements, 
which are more simply expressed and proven in the general theory.
Further, it is much simpler to devise new rely-guarantee refinement laws 
because the algebra gives a rich theory of properties which simplify discovering proofs.

As an example of the way in which the theory generalises rely and guarantee conditions,
in the relational model, as well as being able to express 
a relational rely condition via the process $\FinGuar{r}$,
one can express rely processes, such as the sequence $\FinGuar{r_0} \SSeq \FinGuar{r_1}$,
which cannot be expressed via a relational rely condition.
The closest rely condition is $r_0 \lor r_1$
but that does not represent the fact that the rely transitions from $r_0$ to $r_1$ just once.

\section{Basic commands and refinement}\label{section:basics}

Our presentation separates a core algebraic theory of processes
from an instantiation of that theory as a relational model 
similar to that used by Jones \cite{CoJo07}.
Section~\ref{section:syntax} introduces the operators in our language.
Section~\ref{section:lattices} covers the theory of lattices on which the theory for the language is built.
Section~\ref{section:basic-algebra} gives the algebraic properties of basic commands.
Section~\ref{section:relational-model} gives the relational model 
to provide an intuition for the behaviour of basic commands.

\subsection{Operators and primitive commands}\label{section:syntax}

\begin{figure*}
\figurerule\\
Let $c$ and $d$ be commands, $C$ be a set of commands and $f$ a monotonic function on commands.
The following are the primitive operators and commands used in the algebra.
\begin{eqnarray*}
                       c \nondet d 
                 \Altx c \angelic d  
                 \Altx c \parallel d 
                 \Altx c \together d 
                 \Altx c \quotient d
                 \Altx c \Seq d 
                 \Altx \mu f
                 \Altx \nu f
                 \Altx \Nondet C 
                 \Altx \Angelic C 
                 \Altx \Abort 
                 \Altx \Magic 
                 \Altx \Nil 
                 \Altx \Skip
                 \Altx \Chaos 
\end{eqnarray*}
The precedence of binary operators ranges 
from ``$\nondet$'' on the left having the lowest precedence
to ``$\Seq$'' on the right having the highest precedence,
although ``$\nondet$'' and ``$\angelic$'' have equal precedence.
Unary operators have precedence over binary operators.
The sequential composition $c \Seq d$ is abbreviated as $c \SSeq d$.

\figurerule
\caption{Operators and primitive commands}\label{figure:syntax}
\end{figure*}

The operators and primitive commands of the core language are given in Figure~\ref{figure:syntax}.
Typical 
commands are represented by $c$, $d$, $i$ and $j$;
sets of commands by $C$ and $D$;
and
monotonic functions from commands to commands by $f$.
The language includes non-deterministic choice, 
both binary $(c \nondet d)$ and over a set of commands $(\Nondet C)$, 
which form infima with respect to the refinement ordering,
and their duals $c \angelic d$ and $(\Angelic C)$,
which form suprema.
Additional binary operators are 
parallel composition $(c \parallel d)$, 
sequential composition $(c \Seq d)$,
a \strictconjunction\ operator $(c \together d)$ explained in Section \ref{section:conjunction},
and
a rely quotient operator $(c \quotient d)$ explained in Section~\ref{section:rely}.
Commands include least ($\mu f$) and greatest ($\nu f$) fixed points of monotonic functions over commands.
Primitive commands include:
the top element in the refinement lattice $\Magic$ 
(called \emph{magic} in the refinement calculus);
the bottom element $\Abort$ 
(called \emph{abort});
the command that terminates immediately, $\Nil$,
which is the identity of sequential composition;
the command that does nothing but doesn't constrain its environment, $\Skip$,
which is the identity of parallel composition;
and
the command that can do any non-aborting behaviour, $\Chaos$,
which is the identity of \strictconjunction.

\subsection{Lattices and fixed points}\label{section:lattices}

\begin{figure}
\figurerule\\[1ex]
\begin{minipage}{0.48\textwidth}
\textbf{Lattice}
\begin{eqnarray}
  c_0 \nondet (c_1 \nondet c_2) & = & (c_0 \nondet c_1) \nondet c_2 \labelaxiom{infimum-associative}\\
  c_0 \nondet c_1 & = & c_1 \nondet c_0 \labelaxiom{infimum-commutes} \\
  c \nondet c & = & c \labelaxiom{infimum-idempotent} \\
  c_0 \angelic (c_1 \angelic c_2) & = & (c_0 \angelic c_1) \angelic c_2 \labelaxiom{supremum-associative}\\
  c_0 \angelic c_1 & = & c_1 \angelic c_0 \labelaxiom{supremum-commutes} \\
  c \angelic c & = & c \labelaxiom{supremum-idempotent} \\
  c_0 \nondet (c_0 \angelic c_1) & = & c_0 \labelaxiom{infimum-absorb-supremum} \\
  c_0 \angelic (c_0 \nondet c_1) & = & c_0  \labelaxiom{supremum-absorb-infimum}
\end{eqnarray}
\end{minipage}
\quad
\begin{minipage}{0.48\textwidth}
\textbf{Complete lattice}
\begin{eqnarray}
  c \in C & \implies & \Nondet C \refsto c \labelaxiom{infimum-lower-bound} \\
  (\forall c \in C \spot d \refsto c) & \implies & d \refsto \Nondet C \labelaxiom{infimum-greatest-lower-bound} \\
  c \in C & ~\implies~ & c \refsto \Angelic C \labelaxiom{supremum-upper-bound} \\
  (\forall c \in C \spot c \refsto d) & ~\implies~ & \Angelic C \refsto d \labelaxiom{supremum-least-upper-bound}
\end{eqnarray}
\textbf{Nondeterminism distributes over supremum}
\begin{eqnarray}
  c \nondet (\Angelic D) & = & \Angelic \{ d \in D \spot c \nondet d \} \labelaxiom{infimum-distribute-supremum} 
\end{eqnarray}
\end{minipage}
\begin{minipage}{0.48\textwidth}
\vspace*{2ex}
\textbf{Fixed point axioms}
\begin{eqnarray}
  \mu f & = & f(\mu f)  \labelaxiom{least-fixed-point-unfold} \\
  f(x) \refsto x & \implies & \mu f \refsto x \labelaxiom{least-fixed-point-induction}
\end{eqnarray}
\end{minipage}
\quad
\begin{minipage}{0.48\textwidth}
\begin{eqnarray}
  \nu f & = & f(\nu f)  \labelaxiom{greatest-fixed-point-unfold} \\
  x \refsto f(x) & \implies & x \refsto \nu f  \labelaxiom{greatest-fixed-point-induction}
\end{eqnarray}
\end{minipage}

\vspace*{1ex}
\figurerule

\caption{Axioms for lattices and fixed points}\label{figure:lattices}
\end{figure}

The theory for the language is built on a lattice of commands ordered by \emph{refinement}. 
The refinement relation ``$\refsto$'' is defined in terms of the infimum operator ``$\nondet$'';
refinement is reflexive, anti-symmetric and transitive (a partial order).
\begin{definitionx}[refinement]
For any $c,d$,~~~
\(
  c \refsto d ~\sdefs~ (c \nondet d) = c.
\)
Equivalently $c \refsto d ~\iff~  (c \angelic d) = d$.
\end{definitionx}
The lattice-theoretic axioms of the language are given in Figure~\ref{figure:lattices}.
$\Command$ is the set of all commands and lattice infimum, ``$\nondet$'', corresponds to nondeterministic choice.
\begin{itemize}
\item
$(\Command, \nondet, \angelic)$ forms a lattice with infimum (greatest lower bound) ``$\nondet$'' and 
supremum (least upper bound) ``$\angelic$'',
i.e.\ axioms
(\refaxiom*{infimum-associative}--
\refaxiom*{supremum-absorb-infimum}) hold.
\item
The lattice is \emph{complete},
i.e.\ the infimum $\Nondet C$ and the supremum $\Angelic C$ exist for all sets of commands $C$,
including empty or infinite $C$.
The infima and suprema satisfy axioms 
by (\refaxiom*{infimum-lower-bound}--\refaxiom*{supremum-least-upper-bound}).
\item
The infimum (i.e.\ nondeterministic choice) distributes over arbitrary suprema (\refaxiom*{infimum-distribute-supremum}).
\item
The \ifinv{top}{bottom} element of the lattice is $\Abort$.
It is the identity of ``$\angelic$'' and an annihilator for ``$\nondet$''. \\[-1ex]
\begin{minipage}{0.48\textwidth}
\begin{eqnarray}
  \Abort & \sdefs & \Angelic \{\} = \Nondet \Command \labeldefinition{bottom} 
\end{eqnarray}
\end{minipage}
\begin{minipage}{0.48\textwidth}
\begin{eqnarray}
  c \angelic \Abort & =  c  = & \Abort \angelic c  \labelproperty{supremum-identity} \\
  c \nondet \Abort & = \Abort  = & \Abort \nondet c  \labelproperty{nondet-abort-zero}
\end{eqnarray}
\end{minipage}
\item
The \ifinv{bottom}{top} element of the lattice is $\Magic$.
It is the identity of ``$\nondet$'' and an annihilator for ``$\angelic$''. \\[-1ex]
\begin{minipage}{0.48\textwidth}
\begin{eqnarray}
  \Magic & \sdefs & \Nondet \{\} = \Angelic \Command
\end{eqnarray}
\end{minipage}
\begin{minipage}{0.48\textwidth}
\begin{eqnarray}
  c \nondet \Magic & =  c  = & \Magic \nondet c  \labelproperty{nondet-identity} \\
  c \angelic \Magic & = \Magic  = & \Magic \angelic c  \labelproperty{supremum-magic-zero} 
\end{eqnarray}
\end{minipage}
\end{itemize}
The following law can be used to handle refinement to or from a nondeterministic choice \cite{BackWright98}.
A common special case is if $C$ (or $D$) is a singleton set,
i.e.\ $\Nondet \{c\} = c$ (or $\Nondet \{d\} = d$). 
\begin{lemmax}[non-deterministic-choice]
For any sets $C$ and $D$ over a complete lattice,
\begin{eqnarray*}
  (\forall d \in D \spot (\exists c \in C \spot c \refsto d)) & ~~\implies~~ & (\Nondet C) \refsto (\Nondet D).
\end{eqnarray*}
\end{lemmax}
The reverse implication does not hold in general, e.g. for $C = \{c_0,c_1\}$ and $D = \{c_0 \nondet c_1\}$.

\begin{lemmax}[operator-monotonic]
If a binary operator ``$\circ$'' distributes over non-deterministic choice in both arguments then,
$c_0 \refsto c_1 \land d_0 \refsto d_1 ~\implies~ c_0 \circ d_0 \refsto c_1 \circ d_1$.
\end{lemmax}

For a monotonic function $f$ on a complete lattice,
the least and greatest fixed points of $f$, $\mu f$ and $\nu f$, respectively, satisfy 
axioms (\refaxiom*{least-fixed-point-unfold}-\refaxiom*{greatest-fixed-point-induction}).
As usual, $\mu(\lambda x \spot f(x))$ is abbreviated $\mu x \spot f(x)$.
The following lemma allows reasoning about fixed points \cite{fixedpointcalculus1995,BackWright98}.
\newcommand{\FF}{F}
\newcommand{\GG}{G}
\newcommand{\HH}{H}
\begin{lemmax}[fusion]
For any monotonic functions $\FF$, $\GG$ and $\HH$ on complete lattices with order $\refsto$, 
\begin{eqnarray}
   \FF(\mu \GG) ~\refsto~ \mu \HH  && \mbox{ provided } \FF \circ \GG ~\refsto~ \HH \circ \FF
        \mbox{ and } \FF \mbox{ distributes over arbitrary suprema}
     \labelproperty{fusion-lfp-leq}
\\
   \FF(\mu \GG) ~=~ \mu \HH  && \mbox{ provided } \FF \circ \GG ~=~ \HH \circ \FF
        \mbox{ and } \FF \mbox{ distributes over arbitrary suprema}
     \labelproperty{fusion-lfp-eq}
\\
  \FF(\nu \GG) ~\refines~ \nu \HH    && \mbox{ provided } \FF \circ \GG ~\refines~ \HH \circ \FF
        \mbox{ and } \FF \mbox{ distributes over arbitrary infima}
     \labelproperty{fusion-gfp-geq}
\\
  \FF(\nu \GG) ~=~ \nu \HH    && \mbox{ provided } \FF \circ \GG ~=~ \HH \circ \FF
        \mbox{ and } \FF \mbox{ distributes over arbitrary infima}
     \labelproperty{fusion-gfp-eq}
\end{eqnarray}
where 
$\FF$ distributes over arbitrary suprema if 
$\FF(\Angelic C) = \Angelic \{ c \in C \spot \FF(c) \}$ for all sets of commands $C$,
and
$\FF$ distributes over arbitrary infima if 
$\FF(\Nondet C) = \Nondet \{ c \in C \spot \FF(c) \}$ for all sets of commands $C$.
\end{lemmax}

\subsection{An algebra for concurrency}\label{section:basic-algebra}

The properties of the operators in Figure~\ref{figure:syntax} are given 
in terms of a set of axioms given in \Definition*{concurrent-algebra}.
The axioms have been split into groups which are discussed below.
The main results of the paper depend only on these axioms.
The majority of the axioms are taken from existing algebraic theories of programs 
(such as \cite{Wright04,DBLP:journals/jlp/HoareMSW11}),
the main exceptions being the axioms for \strictconjunction, including the exchange axioms.
The axioms hold for the relational model introduced in Section~\ref{section:relational-model}. 

\begin{figure}
\figurerule\\
The notation $\{ c\in C \spot f \}$ stands for the set of values of the expression $f$ for $c$ an element of $C$. 
\\[2ex]
\begin{minipage}{0.49\textwidth}
\textbf{Sequential}
\begin{eqnarray}
  c_0 \SSeq (c_1 \SSeq c_2) & = & (c_0 \SSeq c_1) \SSeq c_2 \labelaxiom{sequential-associative} \\
  c \SSeq \Nil & = & c   \labelaxiom{sequential-identity-right} \\
  \Nil \SSeq c & = & c   \labelaxiom{sequential-identity-left} \\
  c \SSeq (d_0 \nondet d_1) & = & (c \SSeq d_0) \nondet (c \SSeq d_1) \labelaxiom{sequential-distribute-nondet-left} \\
  (\Nondet C) \SSeq d & = & \Nondet \{ c \in C \spot c \SSeq d \}  \labelaxiom{sequential-distribute-nondet-right} \\
  \Abort \SSeq c & = & c \labelaxiom{sequential-abort-zero-left}
\end{eqnarray}
\textbf{Parallel}
\begin{eqnarray}
  c_0 \parallel (c_1 \parallel c_2) & = & (c_0 \parallel c_1) \parallel c_2 \labelaxiom{parallel-associative}\\
  c_0 \parallel c_1 & = & c_1 \parallel c_0 \labelaxiom{parallel-commutes} \\
  c \parallel \Skip & = & c \labelaxiom{parallel-identity} \\
  (\Nondet C) \parallel d & = & \Nondet \{ c \in C \spot c \parallel d \}  \labelaxiom{parallel-distribute}
\end{eqnarray}
\end{minipage}
\quad
\begin{minipage}{0.48\textwidth}
\textbf{Identities}
\begin{eqnarray}
  \Skip \SSeq \Skip & = & \Skip  \labelaxiom{skip-skip} \\
  \Skip & \refsto & \Nil \labelaxiom{skip-nil}
\end{eqnarray}
\textbf{\Strictconjunction}
\begin{eqnarray}
  c_0 \together (c_1 \together c_2) & = & (c_0 \together c_1) \together c_2  \labelaxiom{conjunction-associative} \\
  c_0 \together c_1 & = & c_1 \together c_0  \labelaxiom{conjunction-commutes} \\
  c \together c & = & c  \labelaxiom{conjunction-idempotent} \\
  c \together \Chaos & = & c  \labelaxiom{conjunction-identity} \\
  \Chaos & \refsto & \Skip  \labelaxiom{chaos-skip} \\
  \Chaos \parallel \Chaos & = & \Chaos  \labelaxiom{chaos-parallel-chaos} \\
  D \neq \emptyset \implies c \together (\Nondet D) & = & \Nondet \{ d \in D \spot c \together d \} \labelaxiom{conjunction-distribute-infimum} \\
  c \together (\Angelic D) & = & \Angelic \{ d \in D \spot c \together d \} \labelaxiom{conjunction-distribute-supremum}
\end{eqnarray}
\end{minipage}
\begin{center}
\begin{minipage}{0.6\textwidth}
\textbf{Weak exchange axioms}
\begin{eqnarray}
  (c_0 \parallel c_1) \together (d_0 \parallel d_1) & ~\refsto~ & (c_0 \together d_0) \parallel (c_1 \together d_1) 
    \labelaxiom{conjunction-exchange-parallel} \\
  (c_0 \SSeq c_1) \together (d_0 \SSeq d_1) & ~\refsto~ & (c_0 \together d_0) \SSeq (c_1 \together d_1) 
    \labelaxiom{conjunction-exchange-sequential}
\end{eqnarray}
\end{minipage}
\end{center}
\figurerule

\caption{Axioms for core language of commands}\label{figure:axioms}
\end{figure}

\begin{definitionx}[concurrent-algebra]
The set of commands $\Command$ satisfies the axioms given in Figure~\ref{figure:axioms}
in addition to the axioms of lattices from Figure~\ref{figure:lattices}.
\end{definitionx}
\begin{itemize}
\item
$(\Command, \Seq\,, \Nil)$ forms a monoid with identity $\Nil$, i.e. axioms (\refaxiom*{sequential-associative}-\refaxiom*{sequential-identity-left}).
Note that the operator ``$\Seq$'' is elided, so that ``$c \Seq d$'' is written ``$c \SSeq d$''. 
\item
Sequential composition distributes over finite non-deterministic choices on the left (\refaxiom*{sequential-distribute-nondet-left}) 
and arbitrary infima on the right (\refaxiom*{sequential-distribute-nondet-right}) and 
and hence it has a left annihilator of  $\Magic$ (\refproperty*{sequential-magic-zero-left});
$\Abort$ is a left annihilator of sequential composition $\Abort$ (\refaxiom*{sequential-abort-zero-left}). 
 \begin{eqnarray}
  \Magic \SSeq c & = & \Magic  \labelproperty{sequential-magic-zero-left}
 \end{eqnarray}
\item
$(\Command, \parallel, \Skip)$ forms a monoid with identity $\Skip$ in which ``$\parallel$'' is commutative, i.e.\ axioms 
(\refaxiom*{parallel-associative}--\refaxiom*{parallel-identity}).
Note that the identity of parallel composition is different to the identity of sequential composition;
that allows a wider range of models, included the relational model introduced in Section~\ref{section:relational-model}.
\item
Parallel distributes over non-deterministic choice of any set of commands
(\refaxiom*{parallel-distribute}),
and hence 
has an annihilator of $\Magic$.
\begin{eqnarray}
  \Magic \parallel c & = & \Magic  \labelproperty{parallel-top-zero}
\end{eqnarray}
\item
The identity of parallel composition, $\Skip$, sequentially composed with itself is equivalent to $\Skip$ (\refaxiom*{skip-skip})
and is refined by the identity of sequential composition, $\Nil$ (\refaxiom*{skip-nil}).

\item
$(\Command, \together, \Chaos)$ forms a monoid with identity $\Chaos$ in which ``$\together$'' is commutative and idempotent, 
i.e.\ axioms (\refaxiom*{conjunction-associative}--\refaxiom*{conjunction-identity}).
\item
$\Chaos$ allows any non-aborting behaviour including $\Skip$ (\refaxiom*{chaos-skip})
and $\Chaos$ in parallel with itself doesn't make it any more (or less) chaotic (\refaxiom*{chaos-parallel-chaos}).
\item
\Strictconjunction\ distributes over the non-deterministic choice of non-empty sets of commands by \refaxiom{conjunction-distribute-infimum}
and hence it distributes over binary choices.
\begin{eqnarray}
  c \together (d_0 \nondet d_1) & = &  (c \together d_0) \nondet (c \together d_1) \labelproperty{conjunction-distribute-infimum}
\end{eqnarray}
\item
\Strictconjunction\ distributes over arbitrary suprema \refaxiom{conjunction-distribute-supremum}
and hence
it has an annihilator of $\Abort$.
\begin{eqnarray}
  c \together \Abort = & \Abort &  = \Abort \together c  \labelproperty{conjunction-abort-zero}
\end{eqnarray}
\item
\Strictconjunction\ does \emph{not} distribute through either parallel or sequential composition,
instead it satisfies the weak exchange axioms 
(\refaxiom*{conjunction-exchange-parallel}) and (\refaxiom*{conjunction-exchange-sequential}).
Note that \refaxiom{conjunction-exchange-parallel} is a refinement rather than an equality 
because, on the left, behaviour of $c_0$ may synchronise with 
behaviour of either $d_0$ or $d_1$,
whereas, on the right, it can only synchronise with behaviour of $d_0$;
\refaxiom{conjunction-exchange-sequential} is similar;
see Section~\ref{section:conjunction} for more details.
\end{itemize}
Note that the set of all commands that refine $\Chaos$ forms a sub-lattice of all non-aborting commands.

The iteration operators are based on von Wright's refinement algebra \cite{Wright04}.
Kleene algebra provides the finite iteration operator $\itkleene{c}$,
which iterates $c$ zero or more times
but only a finite number of times \cite{conway71,Blikle78,kozen97kleene}.
A generalisation of this more appropriate for modelling programs is
the iteration operator, $\itomega{c}$, 
that iterates $c$ zero or more times, 
including the possibility of an infinite number of iterations \cite{Wright04}.
For both these operators the number of iterations they take is 
non-deterministic.
\begin{definitionx}[iteration]
The iteration operators are defined via least ($\mu$) and greatest ($\nu$) 
fixed point operators. 
\\[-1ex]
\begin{minipage}{0.5\textwidth}
\begin{eqnarray}
  \itkleene{c} & ~\sdefs~ & (\finiterec x \spot \Nil ~\nondet~ c \SSeq x)
    \labeldefinition{kleene-iteration-skip} 
\end{eqnarray}
\end{minipage}
\begin{minipage}{0.49\textwidth}
\begin{eqnarray}
  \itomega{c}  & \sdefs & (\infiniterec x \spot \Nil ~\nondet~ c \SSeq x) 
    \labeldefinition{omega-iteration-skip} 
\end{eqnarray}
\end{minipage}
\end{definitionx}
The iteration operators have corresponding induction and folding/unfolding lemmas
\cite{BackWright98,BackWright99,Wright04}.
\begin{lemmax}[fold/unfold]
The iteration unfolding properties follow from fixed point unfolding (\refaxiom*{greatest-fixed-point-unfold}) 
and (\refaxiom*{least-fixed-point-unfold}).
\\[-1ex]
\begin{minipage}{0.5\textwidth}
\begin{eqnarray}
  \itkleene{c} & ~=~ & \Nil ~\nondet~ c \SSeq \itkleene{c} 
    \labelproperty{kleene-unfold}
\end{eqnarray}
\end{minipage}
\begin{minipage}{0.49\textwidth}
\begin{eqnarray}
  \itomega{c}  & = & \Nil ~\nondet~ c \SSeq \itomega{c}
    \labelproperty{omega-unfold} 
\end{eqnarray}
\end{minipage}
\end{lemmax}

\begin{lemmax}[induction]
The iteration induction properties follow from \Lemma{fusion} and 
fixed point induction (\refaxiom*{greatest-fixed-point-induction})  and (\refaxiom*{least-fixed-point-induction}).
\\[-1ex]
\begin{minipage}{0.5\textwidth}
\begin{eqnarray}
  x ~\refsto~  d ~\nondet~ c \SSeq x & ~~\implies~~ & x ~\refsto~ \itkleene{c} \SSeq d
    \labelproperty{kleene-induction}
\end{eqnarray}
\end{minipage}
\begin{minipage}{0.49\textwidth}
\begin{eqnarray}
  d ~\nondet~ c \SSeq x ~\refsto~ x & ~~\implies~~ & \itomega{c} \SSeq d ~\refsto~ x 
    \labelproperty{omega-induction} 
\end{eqnarray}
\end{minipage}
\end{lemmax}
We use the term 
``law'' for theorems about our new operators
and 
``lemma'' for existing theorems from standard theory.
Laws and lemmas share their numbering sequence.
\begin{lawx}[monotonic]
If $c \refsto d$ and $c_0 \refsto d_0$ and $c_1 \refsto d_1$,
all of the following hold. \\[-1ex]
\begin{minipage}{0.5\textwidth}
\begin{eqnarray}
  c_0 \nondet c_1 & ~\refsto~ & d_0 \nondet d_1 
    \labelproperty{nondet-monotonic} \\
  c_0 \parallel c_1 & \refsto & d_0 \parallel d_1
    \labelproperty{parallel-monotonic} \\
  c_0 \SSeq c_1 & \refsto & d_0 \SSeq d_1
    \labelproperty{sequential-monotonic}
\end{eqnarray}
\end{minipage}
\begin{minipage}{0.49\textwidth}
\begin{eqnarray}
  c_0 \together c_1 & \refsto & d_0 \together d_1
    \labelproperty{conjunction-monotonic} \\
  c\FinIter & \refsto & d\FinIter  
    \labelproperty{kleene-monotonic} \\
  c\FinOrInfIter & \refsto & d\FinOrInfIter 
    \labelproperty{omega-monotonic}
\end{eqnarray}
\end{minipage}
\end{lawx}

\begin{proof}
Property (\refproperty*{nondet-monotonic}) holds because non-deterministic choice is associative, commutative and idempotent.
The proofs of (\refproperty*{parallel-monotonic}--\refproperty*{conjunction-monotonic}) follow from \Lemma{operator-monotonic} 
because ``$\Seq$'', ``$\parallel$'' and ``$\together$'' distribute non-deterministic choice in both their 
left 
and 
right 
arguments.
Properties (\refproperty*{kleene-monotonic}) and (\refproperty*{omega-monotonic})
can be shown by induction, respectively, (\refproperty*{kleene-induction}) and (\refproperty*{omega-induction}),
using (\refproperty*{kleene-unfold}) and (\refproperty*{omega-unfold}) (see \cite{Wright04}).
\end{proof}

\subsection{A relational model}\label{section:relational-model}

In this paper we focus on the algebraic laws satisfied by commands 
but it is useful to have a model to gain intuitions and ensure the algebra 
is consistent.
The model used corresponds to the rely-guarantee theory of Jones based on 
Aczel traces \cite{Aczel83,BoerHannemanDeRoever99,DeRoever01,HayesJonesColvin14TR}.
Typical
single-state predicates are represented by $p$
and
binary relations on states by $g$, $q$ and $r$.
The additional commands in the relational model are 
\begin{eqnarray*}
    \cpstep{r} \Altx \cestep{r} \Altx \cgd{p} \Altx  \Pre{p} \Altx  \atomicrel{q} \Altx \Spec{}{}{q} ~. 
\end{eqnarray*}
This set of commands is left open and may be extended with other commands,
for example, tests, assignments, conditionals and loops are added in \cite{HayesJonesColvin14TR}.

States ($\Sigma$) are modelled by a mapping from variable names to values.
The set of program states $\Sigma_{\botstate}$ is extended to include the undefined state $\botstate$,
which is used to denote that the process has aborted.%
\footnote{The symbol $\Abort$ is overloaded between the undefined state and 
the bottom of the lattice of commands, which corresponds to the aborted process.
As usual their meaning is resolved by context.}
An \emph{Aczel trace} consists of an initial state $\sigma \in \Sigma$ and a sequence of steps,
each of which is either a program step labelled $\pstepl{\sigma'}$ 
or an environment step labelled $\estepl{\sigma'}$,
where $\sigma' \in \Sigma_{\botstate}$ is the program state after the step.
In this paper the term ``step'' always means an atomic step
(either of a program or its environment).
A \emph{terminating} Aczel trace ends with a step labelled $\termd$.
The step $\pstepl{\botstate}$ is an aborting step of the program and
the step $\estepl{\botstate}$ allows an aborting step by the environment.
The special steps $\termd$, $\pstepl{\botstate}$ and $\estepl{\botstate}$ can appear only as 
the last step of a (finite) trace.
The set $\Trace$ is the set of all valid Aczel traces.
The notation $[v_1,v_2, \ldots]$ stands for the sequence containing $v_1, v_2, \ldots$.

A set of traces $T$ is \emph{prefix closed} if $(\sigma, [~]) \in T$ for all $\sigma \in \Sigma$
and whenever $(\sigma,t) \in T$ and $t'$ is a prefix of $t$, $(\sigma,t') \in T$.
A set of traces $T$ is \emph{abort closed} if whenever $(\sigma,t \cat [\pstepl{\botstate}]) \in T$,
then for any valid trace $(\sigma, t \cat t') \in \Trace$, $(\sigma, t \cat t') \in T$.
The set of all commands, $\Command$, consists of all the prefix and abort closed subsets of $\Trace$. 

The command
$\cpstep{r}$ performs a single program step with its before and after states related by $r$ and terminates (\refproperty*{semantics-program-step}),
$\cestep{r}$ is similar but performs an environment step (\refproperty*{semantics-env-step}),
$\cestepbot{r}$ represents an environment step that satisfies $r$ or allows a parallel process to abort (\refproperty*{semantics-env-abort}),
$\cgd{p}$ terminates from states satisfying $p$ only (\refproperty*{semantics-guard}),
$\Abort$ aborts immediately and hence can do any behaviour whatsoever (\refproperty*{semantics-abort}),
$\Magic$ can make no steps whatsoever (\refproperty*{semantics-magic}),
and
$\Nil$ terminates immediately from any state (\refproperty*{nil}).
Recall that $\{ x \in S \spot e \}$ stands for the set of values of $e$ for all values of $x$ in the set $S$.\\
\begin{minipage}{0.55\textwidth}
\begin{eqnarray}
  \cpstep{r} & = & \PrefixClose{ \{ (\sigma,\sigma') \in r \spot (\sigma,[\pstepl{\sigma'},\termd] \} }    \labelproperty{semantics-program-step} \\
  \cestep{r} & = & \PrefixClose{ \{ (\sigma,\sigma') \in r \spot (\sigma,[\estepl{\sigma'},\termd] \} }    \labelproperty{semantics-env-step} \\
  \cestepbot{r} & = & \cestep{r} \cup \PrefixClose{ \{ \sigma \in \Sigma \spot (\sigma, [\estepl{\botstate}]) \} } \labelproperty{semantics-env-abort} \\
  \cgd{p} & = & \PrefixClose{ \{ \sigma \in p \spot (\sigma,[\termd] \} }    \labelproperty{semantics-guard} 
\end{eqnarray}
\end{minipage}
\begin{minipage}{0.44\textwidth}
\begin{eqnarray}
  \Abort  & ~=~ & \Trace   \labelproperty{semantics-abort} \\
  \Magic & = & \{ \sigma \in \Sigma \spot (\sigma, [~])\}    \labelproperty{semantics-magic} \\
  \Nil & ~=~ & \cgd{\Sigma}   \labelproperty{nil}  
\end{eqnarray}
\end{minipage}
\\[1ex]
The set of traces of a non-deterministic choice $\Nondet C$ is the union 
$\bigcup C$ and the supremum $\Angelic C$ is the intersection $\bigcap C$.
A trace of a sequential composition $(c \SSeq d)$ is 
any unterminated trace of $c$
or
a terminating trace $t$ of $c$ (minus the $\termd$ step) followed by a trace of $d$
that starts in the final state of $t$.
Note that an unterminated trace may be infinite or it may be a finite trace that does not end in $\termd$.

The traces of $c \parallel d$ are formed by matching traces of $c$ and $d$.
A program step $sc$ of $c$ matches an environment step $sd$ of $d$ 
if their states are the same,
in which case the program step is the step taken by their parallel composition.
Identical environment steps of both $c$ and $d$ match
to give an environment step of their parallel composition.
The following predicate defines matching a step $sc$ of $c$ 
with a step $sd$ of $d$ to give a step $st$ of $c \parallel d$.
\begin{eqnarray*}
  match\_step(sc,sd,st) & ~\sdefs~ & \exists \sigma \in \Sigma_{\botstate} \spot 
      \begin{array}[t]{l}
           sc = \pstepl{\sigma} \land sd = \estepl{\sigma} \land st = \pstepl{\sigma} \lor {} \\
           sc = \estepl{\sigma} \land sd = \pstepl{\sigma} \land st = \pstepl{\sigma} \lor {} \\
           sc = \estepl{\sigma} \land sd = \estepl{\sigma} \land st = \estepl{\sigma} \lor {} \\
           sc = \termd \land sd = \termd \land st = \termd
       \end{array} \\
  match\_trace((\sigma_c,t_c),(\sigma_d,t_d),(\sigma,t)) & ~\sdefs~ & 
       \begin{array}[t]{l}
           \sigma_c = \sigma_d = \sigma \land dom(t_c) = dom(t_d) = dom(t) \land {} \\
           (\forall i \in dom(t) \spot match\_step(t_c(i),t_d(i),t(i))
        \end{array} \\
  c \parallel d & ~\sdefs~ & abort\_close(\{ t \in \Trace \where \exists tc \in c, td \in d \spot match\_trace(tc,td,t) \})
\end{eqnarray*}
Two traces match if they have the same initial state and are the same length 
(including both being infinite) and
all their corresponding steps match.
The parallel composition of $c$ and $d$ consists of all their matching traces.
The abort closure ensures aborting traces can be refined by any other behaviour.

A \strictconjunction\ $c \together d$ represents synchronised step-by-step execution of $c$ and $d$
unless one of them aborts.
Hence 
if both $c$ and $d$ can make a step $\pstepl{\sigma}$ then so can $c \together d$,
if both $c$ and $d$ can make a step $\estepl{\sigma}$ then so can $c \together d$,
if both $c$ and $d$ can make a step $\termd$ then so can $c \together d$,
but 
if either $c$ or $d$ can make an aborting step $\pstepl{\botstate}$ then so can $c \together d$.
The properties of \strictconjunction\ in the relational model are discussed in more detail 
in Section~\ref{section:\strictconjunction-relational}.

Other commands in the relational model are defined as follows,
where $\universalrel$ stands for the universal relation $\Sigma \times \Sigma$ on states.
\vspace*{-2ex}\\
\begin{minipage}[t]{0.5\textwidth}
\begin{eqnarray}
  \Skip & \sdefs & (\cestepbot{\universalrel})\FinOrInfIter  \labelproperty{semantics-skip} \\
  \atomicrel{r} & \sdefs & \Skip \SSeq \cpstep{r} \SSeq \Skip  \labelproperty{semantics-atomicrel} 
\end{eqnarray}
\end{minipage}
\begin{minipage}[t]{0.49\textwidth}
\begin{eqnarray}
  \Pre{p} & \sdefs & \cgd{p} \nondet (\cgd{\lnot p} \SSeq \Abort)  \labelproperty{semantics-precondition} \\
  \envc{r} & \sdefs & (\cpstep{\universalrel} \nondet \cestepbot{r})\FinOrInfIter \SSeq (\Nil \nondet \cestep{\bar{r}} \SSeq \Abort) \labelproperty{semantics-env}
\end{eqnarray}
\end{minipage}\vspace*{1ex}\\
The command 
$\Skip$ does no program steps but allows its environment to do any steps, including abort.
The atomic step command $\atomicrel{r}$ performs a single program step satisfying $r$ (if possible) and allows its environment to do any steps.
The precondition command $\Pre{p}$ characterises an assumption about the initial state --- 
it terminates immediately if the initial state satisfies $p$, otherwise it aborts immediately.
The command $\envc{r}$ characterises an assumption that all steps of its environment satisfy the relation $r$;
it aborts if its environment performs a step that does not satisfy $r$.
The relational commands satisfy the following laws \cite{HayesJonesColvin14TR}.\\
\begin{minipage}[t]{0.5\textwidth}
\begin{eqnarray}
  p_0 \relcontained p_1 & ~\iff~ & \Pre{p_0} \refsto \Pre{p_1} 
    \labelproperty{weaken-precondition} \\
  r_0 \relcontained r_1 & \iff & \envc{r_0} \refsto \envc{r_1}
    \labelproperty{weaken-environment}
\end{eqnarray}
\end{minipage}
\begin{minipage}[t]{0.49\textwidth}
\begin{eqnarray}
  q_1 \relcontained q_0 & ~\iff~ & \atomicrel{q_0} \refsto \atomicrel{q_1}
    \labelproperty{strengthen-postcondition-atomic} 
\end{eqnarray}
\end{minipage}
\vspace{1ex}

Whereas $\Nil$ terminates immediately allowing no program or environment steps,
$\Skip$ allows any number of environment steps, including allowing the environment to abort.
That ensures that $c \parallel \Skip = c$ because 
any trace $tc$ of program, environment or termination steps of $c$ is matched by 
a trace of $\Skip$ to give the same trace $tc$.
Note that $c \parallel \Nil$ either terminates immediately if $c$ can,
otherwise the trace becomes infeasible.
Because $\Nil$ terminates immediately with no intervening environment steps,
$\Pre{p} \SSeq \Nil \SSeq \Pre{p} = \Pre{p}$,
but if $\Nil$ is replaced by $\Skip$,
environment steps allowed by $\Skip$ may change the state thus invalidating $p$
and hence $\Pre{p} \SSeq \Skip \SSeq \Pre{p} = \Pre{p}$ does not hold in general.

\section{\Strictconjunction}\label{section:conjunction}

A \strictconjunction\ of commands $c \together d$ behaves as both $c$ and $d$
provided neither aborts but aborts as soon as either $c$ or $d$ aborts.
If neither process aborts, $c \together d$ is the same as their supremum $c \angelic d$
(which in the relational model forms the intersection of traces).
\Strictconjunction\ was introduced as part of a relational model in \cite{HayesJonesColvin14TR}
but here it is viewed more abstractly via its axioms in \Definition{concurrent-algebra}.
In Section~\ref{section:conjunction-algebra} a set of laws based only on the axioms
of \strictconjunction\ are derived.
\Strictconjunction\  in the relational model is examined in Section~\ref{section:\strictconjunction-relational},
while Section~\ref{section:guarantee-jones} looks at its use for representing relational guarantees
and Section~\ref{section:guarantee-laws} presents a set of laws about relational guarantees.

\subsection{Laws for \strictconjunction}\label{section:conjunction-algebra}

This section presents a number of laws about \strictconjunction\ 
that can be derived from the axioms presented in Section~\ref{section:basic-algebra}.

\begin{lawx}[refine-conjunction]
If $c_0 \refsto d$ and $c_1 \refsto d$,~~~
\(
  c_0 \together c_1 ~\refsto~ d~.
\)
\end{lawx}

\begin{proof}
The proof follows by \Law{monotonic} part (\refproperty*{conjunction-monotonic}) and 
because \strictconjunction\ is idempotent (\refaxiom*{conjunction-idempotent}): \\
\(
    c_0 \together c_1
  ~\refsto~
    d \together d
  ~=~
    d.
\)
\end{proof}

\begin{lawx}[refine-to-conjunction]
If $c \refsto d_0$ and $c \refsto d_1$,~~~
\(
  c ~\refsto~ d_0 \together d_1~.
\)
\end{lawx}

\begin{proof}
The proof follows because \strictconjunction\ is idempotent (\refaxiom*{conjunction-idempotent})
and by \Law{monotonic} part (\refproperty*{conjunction-monotonic}): \\
\(
  c ~=~ c \together c ~\refsto~ d_0 \together d_1~.
\)
\end{proof}
It is \emph{not} the case that $c ~\refsto~ c \together d$ in general, e.g.\ take $d$ to be $\Abort$,
however, if $d$ refines the identity of \strictconjunction, $\Chaos$, it does hold.

\begin{lawx}[conjoin-non-aborting]
If $\Chaos \refsto d$,~~~
\(
  c ~\refsto~ c \together d~.
\)
\end{lawx}

\begin{proof}
The proof follows because $\Chaos$ is the identity of \strictconjunction\ (\refaxiom*{conjunction-identity})
and by \Law{monotonic} part (\refproperty*{conjunction-monotonic}):~~
\(
  c ~=~ c \together \Chaos ~\refsto~ c \together d~.
\)
\end{proof}
The following two laws highlight the difference between ``$\together$'' and ``$\angelic$''.
In general, $c \together d \refsto c \angelic d$ 
but they coincide if both arguments are non-aborting.

\begin{lawx}[conjunction-supremum]
$c \together d \refsto c \angelic d$.
\end{lawx}

\begin{proof}
By \refaxiom{supremum-upper-bound}, both $c \refsto c \angelic d$ and $d \refsto c \angelic d$,
and hence by \Law{refine-conjunction},
\(
  c \together d \refsto c \angelic d~.
\)
\end{proof}

\begin{lawx}[conjunction-supremum-nonaborting]
If $ \Chaos \refsto c$ and $\Chaos \refsto d$,~~~~$c \together d ~=~ c \angelic d$.
\end{lawx}

\begin{proof}
By \Law{conjunction-supremum} $c \together d ~\refsto~ c \angelic d$.
By \Law{conjoin-non-aborting} because both $c$ and $d$ refine $\Chaos$, 
both $c ~\refsto~ c \together d$ and $d ~\refsto c \together d$,
and hence by \refaxiom{supremum-least-upper-bound}, $c \angelic d ~\refsto~ c \together d$.
\end{proof}

\begin{lawx}[conjunction-distribute]
\begin{eqnarray}
  c \together (d_0 \together d_1) & ~=~ & (c \together d_0) \together (c \together d_1)
    \labelproperty{conjunction-distribute-conjunction} \\
  c \together (d_0 \parallel d_1) & \refsto & 
                                                                     (c \together d_0) \parallel (c \together d_1) 
                                                                     \mbox{~~~~~~~~if $c \refsto c \parallel c$} 
    \labelproperty{conjunction-distribute-parallel} \\
  c \together (d_0 \SSeq d_1) & \refsto &
                                                                (c \together d_0) \SSeq (c \together d_1) 
                                                                \mbox{~~~~~~~~~~~\,if $c \refsto c \SSeq c$}
    \labelproperty{conjunction-distribute-sequential} \\
  c\FinIter \together d\FinIter & \refsto &
                                                (c \together d)\FinIter 
    \labelproperty{conjunction-distribute-kleene} \\ 
  c\FinOrInfIter \together d\FinOrInfIter & \refsto &
                                                             (c \together d)\FinOrInfIter 
    \labelproperty{conjunction-distribute-omega} 
\end{eqnarray}
\end{lawx}

\begin{proof}
Property (\refproperty*{conjunction-distribute-conjunction}) follows 
because \strictconjunction\ is idempotent (\refaxiom*{conjunction-idempotent}), 
commutative (\refaxiom*{conjunction-commutes}) 
and associative (\refaxiom*{conjunction-associative}).
For (\refproperty*{conjunction-distribute-parallel}), assuming $c \refsto c \parallel c$,
\begin{refine}
    c \together (d_0 \parallel d_1)
  \Refsto*[by \Law{monotonic} part (\refproperty*{conjunction-monotonic}) assuming $c \refsto c \parallel c$]
    (c \parallel c) \together (d_0 \parallel d_1)
  \Refsto*[exchanging \strictconjunction\ and parallel by \refaxiom{conjunction-exchange-parallel}]
    (c \together d_0) \parallel (c \together d_1)
\end{refine}%
and for (\refproperty*{conjunction-distribute-sequential}), assuming $c \refsto c \SSeq c$,
\begin{refine}
    c \together (d_0 \SSeq d_1)
  \Refsto*[by \Law{monotonic} part (\refproperty*{conjunction-monotonic}) assuming $c \refsto c \SSeq c$]
    (c \SSeq c) \together (d_0 \SSeq d_1)
  \Refsto*[exchanging \strictconjunction\ and sequential by \refaxiom{conjunction-exchange-sequential}]
    (c \together d_0) \SSeq (c \together d_1)
\end{refine}%
Property (\refproperty*{conjunction-distribute-kleene}) holds by
\Lemma{induction} for finite iteration (\refproperty*{kleene-induction}), if
\begin{eqnarray*}
  c\FinIter \together d\FinIter & ~\refsto~ & \Nil \nondet (c \together d) \SSeq (c\FinIter \together d\FinIter),
\end{eqnarray*}
which can be shown as follows.
\begin{refine}
    c\FinIter \together d\FinIter
  \Equals*[by \Lemma{fold/unfold} part (\refproperty*{kleene-unfold})]
    (\Nil \nondet c \SSeq c\FinIter) \together d\FinIter
  \Refsto*[as \strictconjunction\ distributes over non-deterministic choice (\refaxiom*{conjunction-distribute-infimum})]
    (\Nil \together d\FinIter) \nondet (c \SSeq c\FinIter \together d\FinIter)
  \Refsto*[by \Law{refine-conjunction} as by (\refproperty*{kleene-unfold}) $d\FinIter = \Nil \nondet d \SSeq d\FinIter$ and hence $d\FinIter \refsto \Nil$ and $d\FinIter \refsto d \SSeq d\FinIter$]
    \Nil ~\nondet~ (c \SSeq c\FinIter \together d \SSeq d\FinIter)
  \Refsto*[exchanging \strictconjunction\ and sequential by \refaxiom{conjunction-exchange-sequential}]
    \Nil ~\nondet~ (c \together d) \SSeq (c\FinIter \together d\FinIter)
\end{refine}%
For (\refproperty*{conjunction-distribute-omega}) the proof uses \Lemma{fusion} part (\refproperty*{fusion-lfp-leq})
with
function $\FF = (\lambda x \spot c\FinOrInfIter \together x)$,
$\GG = (\lambda x \spot \Nil \nondet d \SSeq x)$ 
and hence $\infiniterec \GG = d\FinOrInfIter$,
and $\HH = (\lambda x \spot \Nil \nondet (c \together d) \SSeq x)$
and hence $\infiniterec \HH = (c \together d)\FinOrInfIter$.
$\FF$, $\GG$ and $\HH$ are monotonic because ``$\nondet$'', ``$\Seq$'' and ``$\together$'' are.
Property (\refproperty*{conjunction-distribute-omega})
corresponds to $\FF(\infiniterec \GG) \refsto \infiniterec \HH$,
and \Lemma{fusion} states that this holds if $\FF \circ \GG \refsto \HH \circ \FF$, 
i.e. for any $x$,
\begin{eqnarray}
  c\FinOrInfIter \together (\Nil \nondet d \SSeq x)   & ~\refsto~ & \Nil \nondet (c \together d) \SSeq (c\FinOrInfIter \together x)
    \labelproperty{fusion-property}
\end{eqnarray}
which holds as follows.
\begin{refine}
    c\FinOrInfIter \together (\Nil \nondet d \SSeq x)
  \Equals*[distributing conjunction over non\-deterministic choice (\refaxiom*{conjunction-distribute-infimum})]
    (c\FinOrInfIter \together \Nil) \nondet (c\FinOrInfIter \together d \SSeq x)
  \Refsto*[by \Law{refine-conjunction} as by (\refproperty*{omega-unfold}) $c\FinOrInfIter = \Nil \nondet c \SSeq c\FinOrInfIter$ and hence $c\FinOrInfIter \refsto \Nil$ and $c\FinOrInfIter \refsto c \SSeq c\FinOrInfIter$]
    \Nil \nondet (c \SSeq c\FinOrInfIter \together d \SSeq x)
  \Refsto*[exchanging \strictconjunction\ and sequential by \refaxiom{conjunction-exchange-sequential}]
    \Nil ~\nondet~ (c \together d) \SSeq  (c\FinOrInfIter \together x)
\end{refine}%
\Lemma{fusion} also requires that $\FF$ distributes over arbitrary \ifinv{infima}{suprema}, 
which holds because \strictconjunction\ distributes over arbitrary \ifinv{infima}{suprema}
(\refaxiom*{conjunction-distribute-supremum}).
\end{proof}

The iterations $c\FinIter$ and $c\FinOrInfIter$ iterating zero times, are equivalent to $\Nil$,
which in the relational model allows no steps at all, not even environment steps,
but for use in guarantees, zero iterations should allow environment steps and hence
the iteration operators $c\FinSkipIter$ and $c\InfSkipIter$ are introduced.
\begin{definition}[guarantee-iteration]\mbox{}\vspace*{-1ex}\\
\begin{minipage}{0.5\textwidth}
\begin{eqnarray}
  c\FinSkipIter & \sdefs & c\FinIter \SSeq \Skip
\end{eqnarray}
\end{minipage}
\begin{minipage}{0.495\textwidth}
\begin{eqnarray}
  c\InfSkipIter & \sdefs & c\FinOrInfIter \SSeq \Skip
\end{eqnarray}
\end{minipage}
\end{definition}
\begin{lemmax}[iteration] 
The following properties follow from \Lemma{fold/unfold} and \Lemma{induction}.
\vspace*{-2ex}\\
\begin{minipage}[t]{0.5\textwidth}
\begin{eqnarray}
  c\InfSkipIter & \refsto & c\FinSkipIter \labelproperty{omega-to-kleene} \\
  c\InfSkipIter & \refsto & \Skip  \labelproperty{omega-skip} 
\end{eqnarray}
\end{minipage}
\begin{minipage}[t]{0.495\textwidth}
\begin{eqnarray}
  c\InfSkipIter & \refsto & c\InfSkipIter \SSeq c\InfSkipIter ~~~~~\mbox{if $c \refsto \Skip \SSeq c$}   \labelproperty{sequential-refines-omega} \\
  c\InfSkipIter & \refsto & (c\InfSkipIter)\FinIter                    ~~~~~\mbox{if $c \refsto \Skip \SSeq c$} \labelproperty{omega-refsto-omega-kleene}
\end{eqnarray}
\end{minipage}
\end{lemmax}

\begin{lawx}[conjunction-distribute-guarantee]
If $c \refsto \Skip \SSeq c$,
\begin{eqnarray}
  c\InfSkipIter \together d\FinOrInfIter & \refsto & (c\InfSkipIter \together d)\FinOrInfIter 
    \labelproperty{conjunction-distribute-omega-omega}
\end{eqnarray}
\end{lawx}

\begin{proof}
The proof 
can be shown using \Lemma{fusion} part (\refproperty*{fusion-lfp-leq}) with 
$\GG = (\lambda x \spot \Nil \nondet d \SSeq x)$ 
and hence $\mu \GG = d\FinOrInfIter$,
$\HH = (\lambda x \spot \Nil \nondet (c\InfSkipIter \together d) \SSeq x)$
and hence $\mu \HH = (c\InfSkipIter \together d)\FinOrInfIter$,
and
$\FF = (\lambda x \spot c\InfSkipIter \together x)$
and hence $\FF(\mu \GG) = c\InfSkipIter \together d\FinOrInfIter$.
Note that $\FF$ distributes over arbitrary suprema
because \strictconjunction\ distributes over arbitrary \ifinv{infima}{suprema}
(\refaxiom*{conjunction-distribute-supremum}).
The proviso for \Lemma{fusion} part (\refproperty*{fusion-lfp-leq}) requires
$c\InfSkipIter \together (\Nil \nondet d \SSeq x) ~\refsto~ 
\Nil \nondet (c\InfSkipIter \together d) \SSeq (c\InfSkipIter \together x)$
which holds as follows.
\begin{refine}
    c\InfSkipIter \together (\Nil \nondet d \SSeq x)
  \Equals*[distributing \strictconjunction\ over non-deterministic choice (\refaxiom*{conjunction-distribute-infimum})]
    (c\InfSkipIter \together \Nil) \nondet (c\InfSkipIter \together d \SSeq x)
  \Refsto*[by \Law{refine-conjunction} as $c\InfSkipIter \refsto \Skip \refsto \Nil$ by (\refproperty*{omega-skip}) and (\refaxiom*{skip-nil}) and $c\InfSkipIter \refsto c\InfSkipIter \SSeq c\InfSkipIter$ by (\refproperty*{sequential-refines-omega}) as $c \refsto \Skip \SSeq c$]
    \Nil \nondet (c\InfSkipIter \SSeq c\InfSkipIter \together d \SSeq x)
  \Refsto*[exchanging \strictconjunction\ and sequential composition by \refaxiom{conjunction-exchange-sequential}]
    \Nil \nondet (c\InfSkipIter \together d) \SSeq (c\InfSkipIter \together x)
\end{refine}%
\end{proof}

\subsection{\Strictconjunction\ in the relational model}\label{section:\strictconjunction-relational}

In the relational model \strictconjunction\ corresponds to synchronised execution of 
atomic steps by both processes unless either process aborts,
i.e.\ every non-aborting step taken by $c \together d$ must be a step allowed by both $c$ and $d$.
If either process aborts, the conjunction aborts (\refproperty*{conjunction-abort-zero}).
The \strictconjunction\ of two atomic step commands $\atomicrel{g}$ and $\atomicrel{r}$
can perform a program step that satisfies both $g$ and $r$
(\refproperty*{atomic-conjoin-atomic}).
An atomic step $\atomicrel{g}$ allows any environment step whatsoever 
and hence two atomic step commands synchronise trivially on environment steps.
More generally, the first program steps of conjoined commands synchronise
followed by the \strictconjunction\ of the remainder of both commands
(\refproperty*{non-skip-conjoin-non-skip}).
If one command in a \strictconjunction\ must do a program step but the other cannot, 
their conjunction never terminates and does no program steps
(\refproperty*{skip-conjoin-non-skip}).

\noindent
\begin{minipage}[t]{0.5\textwidth}
\begin{eqnarray}
  \atomicrel{g} \together \atomicrel{r} & = & \atomicrel{g \relint r}
    \labelproperty{atomic-conjoin-atomic} \\
  (\atomicrel{g} \SSeq c) \together (\atomicrel{r} \SSeq d) & = & \atomicrel{g \relint r} \SSeq (c \together d)
    \labelproperty{non-skip-conjoin-non-skip}
\end{eqnarray}%
\end{minipage}
\begin{minipage}[t]{0.49\textwidth}
\begin{eqnarray}
  \Skip \together (\atomicrel{g} \SSeq c) & = & \Skip \SSeq \Magic 
    \labelproperty{skip-conjoin-non-skip} 
\end{eqnarray}%
\end{minipage}
\vspace{1ex}

The command
$\Chaos$ performs any sequence of non-aborting program steps and allows any environment steps,
while
$\Term$ allows only a finite sequence of non-aborting program steps and any environment steps.
Both are defined in terms of the iteration operators that allow environment steps for zero iterations.
\vspace*{-2ex}
\\
\begin{minipage}[t]{0.5\textwidth}
\begin{eqnarray}
  \Chaos & \sdefs & \InfGuar{\universalrel}  \labelproperty{semantics-chaos} 
\end{eqnarray}%
\end{minipage}
\begin{minipage}[t]{0.49\textwidth}
\begin{eqnarray}
  \Term   & \sdefs & \FinGuar{\universalrel}  \labelproperty{semantics-term} 
\end{eqnarray}
\end{minipage}
\vspace{1ex}
\\
Iterations of atomic steps satisfy the following properties \cite{HayesJonesColvin14TR}.\\[-2ex]
\begin{minipage}[t]{0.5\textwidth}
\begin{eqnarray}
  r_1 \relcontained r_0 & ~\implies~ & \FinGuar{r_0} \refsto \FinGuar{r_1}  
    \labelproperty{strengthen-postcondition-atomic-kleene} \\
  r_1 \relcontained r_0 & \implies & \InfGuar{r_0} \refsto \InfGuar{r_1}  
    \labelproperty{strengthen-postcondition-atomic-omega} 
\end{eqnarray}
\end{minipage}
\begin{minipage}[t]{0.49\textwidth}
\begin{eqnarray}
  \FinGuar{r_0 \cup r_1} & ~=~ &  \FinGuar{r_0} \parallel \FinGuar{r_1}
    \labelproperty{parallel-atomic-kleene} \\
  \InfGuar{r_0 \cup r_1} & \refsto & \InfGuar{r_0} \parallel \InfGuar{r_1}
    \labelproperty{parallel-atomic-omega2} \\
  \InfGuar{r} & = & \InfGuar{r} \parallel \InfGuar{r}
    \labelproperty{parallel-atomic-omega1} 
\end{eqnarray}
\end{minipage}\vspace{1ex}\\
Properties (\refproperty*{strengthen-postcondition-atomic-kleene}) and (\refproperty*{strengthen-postcondition-atomic-omega}) 
follow using (\refproperty*{strengthen-postcondition-atomic}) from 
(\refproperty*{kleene-monotonic}) and (\refproperty*{omega-monotonic}), respectively.

In the relational model 
a command $c$ preconditioned by the state predicate $p$ 
is represented by  
$(\Pre{p} \SSeq c)$.
If $p$ holds initially, $\Pre{p}$ behaves as $\Nil$
and hence $(\Pre{p} \SSeq c)$ behaves as $c$
but if $p$ does not hold initially, the preconditioned command aborts.
A precondition distributes into both a \strictconjunction\ and into a parallel composition.
These laws follow from the definition of a precondition command (\refproperty*{semantics-precondition})
and distribution properties in the relational semantics.
\begin{lawx}[precondition-conjunction]~~~
\(
  \Pre{p} \SSeq ( c \together d) ~~=~~ (\Pre{p} \SSeq c) \together (\Pre{p} \SSeq d)~.
\)
\end{lawx}
\begin{lawx}[precondition-parallel]~~~
\(
  \Pre{p} \SSeq ( c \parallel d) ~~=~~ (\Pre{p} \SSeq c) \parallel (\Pre{p} \SSeq d)~.
\)
\end{lawx}

Morgan's specification command, $\Spec{}{}{q}$, is refined by any program 
that terminates with its initial and final states related by $q$
provided there is no interference from the environment
\cite{TSS}.
\begin{eqnarray}
  \Spec{}{}{q} & ~\sdefs~ & \Nondet \{ \sigma \in \Sigma \spot \cgd{\{\sigma\}} \SSeq \Term \SSeq \cgd{\{\sigma' \where (\sigma,\sigma') \in q \}} \} 
    \together \envc{\id}
\end{eqnarray}
The behaviour of $\Spec{}{}{q}$ consists of terminating traces that start in some state $\sigma$
and terminate in a state $\sigma'$ such that $(\sigma,\sigma') \in q$.
It assumes all steps of its environment do not modify the state (i.e.\ satisfy the identity relation $\id$).
Its behaviour includes finite infeasible traces starting from any state 
and traces ending in an infinite sequence of environment steps.
Conjoining two specifications achieves the conjunction of their postconditions. 
\begin{eqnarray}
  \Spec{}{}{q_0} \together \Spec{}{}{q_1} & ~=~ & \Spec{}{}{q_0 \cap q_1}
    \labelproperty{spec-conjoin-spec}
\end{eqnarray}

\subsection{Relationship to Jones-style guarantee}\label{section:guarantee-jones}

Jones introduced the idea of using a guarantee condition $g$, 
a binary relation between states, 
to express the fact that every atomic program step a process makes is guaranteed 
to satisfy $g$ between its before-state and after-state \cite{jon83a}.
The relation $g$ is required to be reflexive so that stuttering steps are allowed.
A guarantee $g$ on a terminating command $c$ can be defined in terms of 
a \strictconjunction\ as $\FinGuar{g} \together c$.
The \strictconjunction\ with $\FinGuar{g}$ restricts the behaviour  of $c$ so that
every atomic program step satisfies $g$.
The command $\FinGuar{g}$ is used rather than $\atomicrel{g}\FinIter$
so that zero iterations corresponds to $\Skip$ rather than $\Nil$
and hence does not constrain environment steps in this case.
More generally, if $c$ is not assumed to be terminating, 
a guarantee is represented by $\InfGuar{g} \together c$.
Possibly infinite iteration is used rather than finite iteration 
because \strictconjunction\ with finite iteration forces termination
and hence is too strong
\cite{HayesJonesColvin14TR}.
Termination of $\InfGuar{g} \together c$ depends only on 
whether $c$ terminates if its traces are restricted to program steps satisfying $g$.
The guarantee component $\InfGuar{g}$ is non-aborting 
and hence any aborting behaviour can only arise from $c$.
Using the supremum operator $\InfGuar{g} \angelic c$
would be too strong a guarantee because $\InfGuar{g}$ has only non-aborting
traces and hence would mask any aborting behaviour of $c$.

A guarantee relation $g$ in the style of Jones is represented here by 
an iterated atomic step satisfying the relation, either $\FinGuar{g}$ or $\InfGuar{g}$.
By treating guarantees as processes more expressive guarantee conditions can
be expressed, for example, the process $\FinGuar{g_0} \SSeq \FinGuar{g_1}$
represents a guarantee of $g_0$ initially, followed at some point by a switch to a guarantee of $g_1$.
As another example, the process $\FinGuar{\id} \SSeq \atomicrel{g} \SSeq \FinGuar{\id}$
represents a guarantee to perform a single step satisfying $g$ surrounded by any finite number of
steps that don't modify any variables.
Neither of these guarantee processes can be represented as a single guarantee relation
unless additional variables that distinguish the phases of the guarantees are used.
It is possible to encode a sequence such as $\FinGuar{g_0} \SSeq \FinGuar{g_1}$
via the use of an additional boolean variable $b$ which is initially false:
$(\lnot b \land g_0) \lor (b \land g_1 \land b')$,
where it is assumed $b$ is set to true for the transition from a guarantee of $g_0$ to $g_1$.

\subsection{Laws for guarantees}\label{section:guarantee-laws}

If $g_0 \relcontained g_1$, then a guarantee of $g_0$ is stronger than a guarantee of $g_1$.
\begin{lawx}[guarantee-strengthen]
For any command $c$ and relations $g_0$ and $g_1$
such that $g_0 \relcontained g_1$,
\begin{eqnarray*}
  \InfGuar{g_1} \together c & ~\refsto~ & \InfGuar{g_0} \together c~.
\end{eqnarray*}
\end{lawx}

\begin{proof}
By (\refproperty*{strengthen-postcondition-atomic-omega}), 
$\InfGuar{g_1} \refsto \InfGuar{g_0}$,
and hence the law follows by \Law{monotonic} part (\refproperty*{conjunction-monotonic}).
\end{proof}

\begin{lawx}[guarantee-introduce]~~~
\(
  c  ~\refsto~ \InfGuar{g} \together c~.
\)
\end{lawx}
\begin{proof}
The proof follows by \Law{conjoin-non-aborting} because by (\refproperty*{semantics-chaos})
$\Chaos = \InfGuar{\universalrel} \refsto \InfGuar{g}$ by (\refproperty*{strengthen-postcondition-atomic-omega}).
\end{proof}

\begin{lawx}[conjunction-atomic-iterated]~~~
\(
  \InfGuar{g_0} \together \InfGuar{g_1} ~=~ \InfGuar{g_0 \relint g_1}
\)
\end{lawx}

\begin{proof}
The refinement from left to right follows 
by \Law{refine-conjunction} because by (\refproperty*{strengthen-postcondition-atomic-omega})
both $\InfGuar{g_0}$ and $\InfGuar{g_1}$ are refined by $\InfGuar{g_0 \cap g_1}$.
The refinement  from right to left can be proved using \Lemma{induction} part (\refproperty*{omega-induction})
using (\refproperty*{non-skip-conjoin-non-skip}) and (\refproperty*{skip-conjoin-non-skip}).
\end{proof}

\begin{lawx}[guarantee-nested]~~~
\(
  \InfGuar{g_0} \together \InfGuar{g_1} \together c ~~=~~ \InfGuar{g_0 \relint g_1} \together c
\)
\end{lawx}

\begin{proof}
By \Law{conjunction-atomic-iterated}, 
$\InfGuar{g_0} \together \InfGuar{g_1} = \InfGuar{g_0 \relint g_1}$.
\end{proof}

A guarantee distributes through 
non-deterministic choice, \strictconjunction, parallel and sequential composition,
and finite and infinite iterations.
\begin{lawx}[guarantee-distribute]
\begin{eqnarray}
  \InfGuar{g} \together (c \nondet d) & ~=~ & (\InfGuar{g} \together c) \nondet (\InfGuar{g} \together d) 
  \labelproperty{guarantee-distribute-nondet} \\
  \InfGuar{g} \together (c \together d) & = & (\InfGuar{g} \together c) \together (\InfGuar{g} \together d) 
  \labelproperty{guarantee-distribute-conjunction} \\
  \InfGuar{g} \together (c \parallel d) & \refsto & (\InfGuar{g} \together c) \parallel (\InfGuar{g} \together d) 
  \labelproperty{guarantee-distribute-parallel} \\
  \InfGuar{g} \together (c \SSeq d) & \refsto & (\InfGuar{g} \together c) \SSeq (\InfGuar{g} \together d) 
  \labelproperty{guarantee-distribute-sequential} \\
  \InfGuar{g} \together c\FinIter & \refsto & (\InfGuar{g} \together c)\FinIter
  \labelproperty{guarantee-distribute-kleene} \\
  \InfGuar{g} \together c\FinOrInfIter & \refsto & (\InfGuar{g} \together c)\FinOrInfIter
  \labelproperty{guarantee-distribute-omega} 
\end{eqnarray}
\end{lawx}

\begin{proof}
Property (\refproperty*{guarantee-distribute-nondet}) holds because \strictconjunction\ distributes over non-deterministic choice 
(\refaxiom*{conjunction-distribute-infimum}),
and (\refproperty*{guarantee-distribute-conjunction}--\refproperty*{guarantee-distribute-kleene}) 
hold by the corresponding properties (\refproperty*{conjunction-distribute-conjunction}--\refproperty*{conjunction-distribute-kleene}) of \Law{conjunction-distribute}.
For property (\refproperty*{guarantee-distribute-parallel}) the proviso holds 
because $\InfGuar{g} = \InfGuar{g} \parallel \InfGuar{g}$
by (\refproperty*{parallel-atomic-omega1});
and
for property (\refproperty*{guarantee-distribute-sequential}) the proviso holds because
$\InfGuar{g} ~\refsto~ \InfGuar{g} \SSeq \InfGuar{g}$
by (\refproperty*{sequential-refines-omega}).
Property (\refproperty*{guarantee-distribute-kleene}) holds by (\refproperty*{conjunction-distribute-kleene}) 
because $\InfGuar{g} \refsto (\InfGuar{g})\FinIter$ by (\refproperty*{omega-refsto-omega-kleene}).
Both (\refproperty*{sequential-refines-omega}) and (\refproperty*{omega-refsto-omega-kleene})
require the side condition $\atomicrel{g} \refsto \Skip \SSeq \atomicrel{g}$,
which holds by (\refproperty*{semantics-atomicrel}).
Property (\refproperty*{guarantee-distribute-omega}) follows from \Law{conjunction-distribute-guarantee}.
\end{proof}

\section{The rely quotient command}\label{section:rely}

Jones introduced the idea of a rely condition,
a reflexive relation assumed to be satisfied by 
every atomic step of the interference from the environment of a process \cite{jon83a}.
In essence it abstracts the environment by a process $\FinGuar{r}$ 
that executes steps satisfying the rely condition $r$.
In the general algebra the environment is represented by an arbitrary process $i$.
The rules of Jones then become a special case 
when $i = \FinGuar{r}$ (see Section~\ref{section:Jones-rely}).
To handle relies in the general algebra, a rely quotient operator ``$\quotient$'' is introduced.
It is defined so that $c \quotient i$ in parallel with $i$ implements $c$,
i.e.,
\begin{eqnarray}\label{rely-motivation}
  c & ~\refsto~ & (c \quotient i) \parallel i~,
\end{eqnarray}
and furthermore for any process $d$, if $c \refsto d \parallel i$ then $c \quotient i \refsto d$.
For example, because $\FinGuar{r_0 \lor r_1} \refsto \FinGuar{r_0} \parallel \FinGuar{r_1}$
holds in the relational model,
one refinement of the quotient $\FinGuar{r_0 \lor r_1} \quotient \FinGuar{r_1}$
is $\FinGuar{r_0}$.

The motivation for the rely quotient is similar to that for 
the weakest pre- and post-specifications of Hoare and He \cite{HoareHe86},
although they deal with residuals of sequential composition rather than parallel composition,
and
weakest environment of Chaochen and Hoare \cite{ChaochenHoare81,Chaochen:1982}.
The rely quotient $c \quotient i$ is defined as the non-deterministic choice over all commands
$d$ satisfying the defining property of the rely quotient:
$c ~\refsto~ d \parallel i$.
\begin{definitionx}[rely-quotient]~~~
\(
  c \quotient i ~~\sdefs~~ \Nondet \{ d \where (c ~\refsto~ d \parallel i) \}~.
\)
\end{definitionx}
This definition is similar to defining division over the positive integers in terms of multiplication
and minimum ($\Nondet$).
\begin{eqnarray*}
  \lceil c/i \rceil \sdefs \Nondet \{ d \where (c \leq d \times i) \}
\end{eqnarray*}
The only command $d$ satisfying $c ~\refsto~ d \parallel i$ might be the infeasible command $\Magic$,
in which case $c \quotient i$ is infeasible.
In particular, taking the interference $i$ to be the aborting process $\Abort$ gives,
$c \quotient \Abort = \Nondet \{ d \where (c \refsto d \parallel \Abort) \} = \Magic$,
unless $c = \Abort$, in which case $\Abort \quotient \Abort = \Abort$.

Because the rely quotient operation is defined in terms of nondeterministic choice and parallel composition,
its instantiation in the relational model follows directly from its definition.
For completeness, an expansion of its definition in the relational model is given below,
in which $\quotient_{\!r}$ and $\parallel_r$ stand for the interpretations of these operators in the relational model;
recall that nondeterministic choice corresponds to set union and refinement to set containment.
\begin{eqnarray*}
  c \quotient_{\!r} i & = & \bigcup \{ d \in \Command \where c \supseteq d \parallel_r i \} \\
    & = & \bigcup \{ d \in \Command \where c \supseteq abort\_close(\{ t \in \Trace \where \exists td \in d, ti \in i \spot match\_trace(td,ti,t) \}) \} 
\end{eqnarray*}
A full appreciation of the utility of the rely quotient operator flows from its use in 
introducing a parallel composition in Section~\ref{section:parallel}
but first we examine a set of basic laws that it satisfies.

\subsection{Laws for rely quotients}

The following law shows that the rely quotient command satisfies its motivating property (\ref{rely-motivation}).
The law corresponds to $c \leq \lceil c/i \rceil \times i$ for positive integer division.
\begin{lawx}[rely-quotient]~~~
\(
  c ~~\refsto~~ (c \quotient i) ~\parallel~ i~.
\)
\end{lawx}

\begin{proof}
The notation $\{ x \where p \spot e \}$ used below represents the set of values of the expression $e$
for $x$ ranging over values that satisfy the predicate $p$.
\begin{refine}
  c ~~\refsto~~ (c \quotient i) ~\parallel~ i
 \IFF*[by \Definition{rely-quotient}]
  c ~~\refsto~~ \Nondet \{ d \where (c ~\refsto~ d \parallel i) \} \parallel i
 \IFF*[distributing parallel over non-deterministic choice (\refaxiom*{parallel-distribute})]
  c ~~\refsto~~ \Nondet \{ d \where (c ~\refsto~ d \parallel i) \spot (d \parallel i) \}
 \ImpliedBy*[by \Lemma{non-deterministic-choice}]
  \forall d \in \{ d \where (c ~\refsto~ d \parallel i) \} ~\spot~ c \refsto (d \parallel i)
\end{refine}
\end{proof}

The following fundamental law shows that the rely quotient is the least command
satisfying its defining property.
It provides the basis for the proof of many of the
laws that follow 
and
shows the Galois connection between 
rely quotient 
and 
parallel composition 
\cite{Aarts92,Backhouse02}.
It corresponds to $\lceil c/i \rceil \leq d \iff c \leq d * i$ for positive integer division.
\pagebreak[3]

\begin{lawx}[rely-refinement]~~~
\(
   c \quotient i ~\refsto~ d ~~~\iff~~~c ~\refsto~ d \parallel i~.
\)
\end{lawx}

\begin{proof}
For the proof from right to left assume $c ~\refsto~ d \parallel i$.
\begin{refine}
  c \quotient i ~\refsto~ d
 \IFF*[by \Definition{rely-quotient}]
  \Nondet \{ d_1 \where (c ~\refsto~ d_1 \parallel i) \} ~\refsto~ d
 \ImpliedBy*[by \Lemma{non-deterministic-choice}]
  \exists d_0 \in \{ d_1 \where (c ~\refsto~ d_1 \parallel i) \} ~\spot~ d_0 \refsto d
 \ImpliedBy*[by assumption $d \in \{ d_1 \where (c ~\refsto~ d_1 \parallel i) \}$]
  d \refsto d
\end{refine}
The proof from left to right assumes $c \quotient i ~\refsto~ d$ and starts with \Law{rely-quotient}.
\begin{refine}
  c ~~\refsto~~ (c \quotient i) \parallel i
 \Implies*[by \Law{monotonic} part (\refproperty*{parallel-monotonic}) as $c \quotient i \refsto d$]
  c ~~\refsto~~ d \parallel i
\end{refine}%
\end{proof}
The property in \Law{rely-refinement} could be used as an alternative 
definition of the rely quotient operator.
From Galois theory, the rely quotient (lower adjoint) is uniquely defined by
the Galois connection provided parallel distributes over non-deterministic choice (\refaxiom*{parallel-distribute}).

Because $\Skip$ is the identity of parallel composition,
it is also the right identity of the rely quotient.
This is similar to 1 being the right identity of integer division ($c/1 = c$).
\begin{lawx}[rely-identity-right]~~~
\(
  c \quotient \Skip ~=~ c
\)
\end{lawx}

\begin{proof}
The law holds by indirect equality if for all $x$,
$c \quotient \Skip \refsto x \iff c \refsto x$,
which holds by \Law{rely-refinement} as follows:
\(
  c \quotient \Skip \refsto x ~~\iff~~c \refsto x \parallel \Skip ~~\iff~~ c \refsto x. 
\)  
\end{proof}

The following two laws correspond to
$c \leq d \implies \lceil c/i \rceil \leq \lceil d/i \rceil$ 
and
$i \leq j \implies \lceil c/j \rceil \leq \lceil c/i \rceil$
for positive integer division.
\begin{lawx}[rely-monotonic]~~~
\(
  c \refsto d ~~~\implies~~~ (c \quotient i) ~\refsto~ (d \quotient i)~.
\)
\end{lawx}

\begin{proof}
By \Law{rely-refinement}, $(c \quotient i) \refsto (d \quotient i)$ holds
if $c ~\refsto~ (d \quotient i) \parallel i$,
which holds by the assumption $c \refsto d$ and \Law{rely-quotient} because
\(
    c 
  ~\refsto~
    d
  ~\refsto~
    (d \quotient i) \parallel i~.
\)
\end{proof}

\begin{lawx}[rely-weaken]~~~
\(
  i \refsto j ~~~\implies~~~ (c \quotient j) ~\refsto~ (c \quotient i)~.
\)
\end{lawx}

\begin{proof}
By \Law{rely-refinement} $(c \quotient j) ~\refsto~ (c \quotient i)$ holds
if $c ~\refsto~ (c \quotient i) \parallel j$,
which holds as follows.
\begin{refine}
    c 
  \Refsto*[by \Law{rely-quotient}] 
    (c \quotient i) \parallel i 
  \Refsto*[by \Law{monotonic} part (\refproperty*{parallel-monotonic}) as $i \refsto j$]
    (c \quotient i) \parallel j
\end{refine}
\end{proof}

\begin{relational}%
[Italic text between horizontal lines partitions out material that applies only to the relational model.] \\
For relational rely conditions,
if 
$r_1 \relcontained r_0$,
then by (\refproperty*{strengthen-postcondition-atomic-kleene}),
$\FinGuar{r_0} \refsto \FinGuar{r_1}$,
and applying \Law{rely-weaken} gives
$(c \quotient \FinGuar{r_1}) \refsto (c \quotient \FinGuar{r_0})$,
i.e. the relational rely condition can be weakened in a refinement. \\
\end{relational}
A nested rely $(c \quotient j) \quotient i$ corresponds to implementing $c$
within environment $j$, all within in environment $i$,
i.e.\ $c$ is implemented in environment $i \parallel j$.
The next law corresponds to $\lceil \lceil c/i \rceil /j\rceil = \lceil c/(i \times j) \rceil$
for positive integer division.
\begin{lawx}[rely-nested]~~~
\(
  (c \quotient j) \quotient i ~~=~~ c \quotient (i \parallel j)~.
\)
\end{lawx}

\begin{proof}
The law follows by indirect equality if for all $x$, 
$(c \quotient j) \quotient i \refsto x ~\iff~ c \quotient (i \parallel j) \refsto x$,
which is shown as follows.
\begin{refine}
  (c \quotient j) \quotient i \refsto x
 \IFF*[by \Law{rely-refinement}]
  c \quotient j \refsto x \parallel i
 \IFF*[by \Law{rely-refinement}]
  c \refsto x \parallel i \parallel j
 \IFF*[by \Law{rely-refinement}]
  c \quotient (i \parallel j) \refsto x
\end{refine}
\end{proof}
\noindent
Because parallel is commutative, 
it follows that
\(
(c \quotient j) \quotient i ~=~ c \quotient (i \parallel j) ~=~ c \quotient (j \parallel i) ~=~ (c \quotient i) \quotient j.
\)

\begin{relational}%
For relational rely conditions
by property (\refproperty*{parallel-atomic-kleene}), 
\(
  \FinGuar{r_0} \parallel \FinGuar{r_1} ~=~ \FinGuar{r_0 \cup r_1},
\)
and hence by \Law{rely-nested} 
nested relational relies of $r_0$ and $r_1$ give an effective rely of $r_0 \cup r_1$.
\begin{eqnarray}
    (c \quotient \FinGuar{r_1}) \quotient \FinGuar{r_0}  
  ~=~
    c \quotient (\FinGuar{r_0} \parallel \FinGuar{r_1}) 
  ~=~
    c \quotient \FinGuar{r_0 \cup r_1}~.
  \labelproperty{rely-nested-rel}
\end{eqnarray}%
\end{relational}

\section{Parallel-introduction law}\label{section:parallel}

The prime motivation of Jones \cite{jon83a} for introducing rely and guarantee conditions
was to support reasoning about parallel compositions.
In the current paper 
a guarantee condition is generalised to a \strictconjunction\ with a process, 
and
a rely condition by a rely quotient by a process.
\Law{parallel-introduce} provides an general law for introducing a parallel composition.
The guarantee $j$ of the first branch of the parallel corresponds to the rely of the second branch and vice versa for $i$.
\begin{lawx}[parallel-introduce]~~~
\(
  c \together d ~~\refsto~~ (j \together (c \quotient i)) ~\parallel~ (i \together (d \quotient j))
\)
\end{lawx}

\begin{proof}
By  \Law{rely-quotient} both 
\(
  c  ~\refsto~  (c \quotient i) \parallel i 
\)
and
\(
  d  ~\refsto~  (d \quotient j) \parallel j 
\)
and hence the proof follows using 
these two properties
in the first step.
\begin{refine}
  c \together d
 \Refsto*[by \Law{monotonic} part (\refproperty*{conjunction-monotonic}) and parallel is commutative (\refaxiom*{parallel-commutes})]
  ((c \quotient i) \parallel i) ~\together~ (j \parallel (d \quotient j))
 \Refsto*[exchanging \strictconjunction\ and parallel by \refaxiom{conjunction-exchange-parallel}]
  ((c \quotient i) \together j) ~\parallel~ (i \together (d \quotient j))
\end{refine}%
\end{proof}
The simplicity and elegance of the proof of this fundamental law for handling rely-guarantee
concurrency is an indication that \strictconjunction\ and rely quotient are well chosen abstractions.
The relationship to the parallel law of Jones is explored in Section~\ref{section:Jones-rely}
but first distribution properties of rely quotients need to be explored.

\section{Distribution of rely quotients}\label{section:rely-distribution}

\Law{parallel-introduce} introduces rely quotients of the form $c \quotient i$
for some specification $c$.
One way of refining such a quotient is to refine $c$,
for example, $c$ may be refined to a sequential composition $c_0 \SSeq c_1$.
\Law{rely-monotonic} then gives that $c \quotient i \refsto (c_0 \SSeq c_1) \quotient i$.
To further refine this it is useful to have a distribution law that allows the rely quotient
to be distributed over the sequential composition,
i.e.\ $(c_0 \SSeq c_1) \quotient i ~\refsto~(c_0 \quotient i) \SSeq (c_1 \quotient i)$.
A proviso is needed for this refinement to be valid (see \Law*{rely-distribute-sequential} below).
This section investigates laws for distributing rely quotients over the other operators.
A rely quotient distributes straightforwardly over both
\strictconjunction\ and non-deterministic choice.
\begin{lawx}[rely-distribute-conjunction]~~~
\(
  (c \together d) \quotient i ~~\refsto~~ (c \quotient i) \together (d \quotient i)
\)
\end{lawx}

\begin{proof}
By \Law{rely-refinement} the law is equivalent to 
$c \together d ~\refsto~ ((c \quotient i) \together (d \quotient i)) \parallel i $.
\begin{refine}
  c \together d
 \Refsto*[by \Law{rely-quotient} twice]
  ((c \quotient i) \parallel i) ~\together~ ((d \quotient i) \parallel i)
 \Refsto*[exchanging \strictconjunction\ and parallel by \refaxiom{conjunction-exchange-parallel}]
  ((c \quotient i) \together (d \quotient i)) ~\parallel~ (i \together i)
 \Equals*[as ``$\together$'' is idempotent (\refaxiom*{conjunction-idempotent})]
  ((c \quotient i) \together (d \quotient i)) ~\parallel~ i 
\end{refine}%
\end{proof}

\begin{lawx}[rely-distribute-choice]~~~
\(
  (c \nondet d) \quotient i ~~\refsto~~ (c \quotient i) \nondet (d \quotient i)
\)
\end{lawx}

\begin{proof}
By \Law{rely-refinement} the law is equivalent to 
$c \nondet d ~\refsto~ ((c \quotient i) \nondet (d \quotient i)) \parallel i $.
\begin{refine}
  c \nondet d
 \Refsto*[by \Law{rely-quotient} twice]
  ((c \quotient i) \parallel i) ~\nondet~ ((d \quotient i) \parallel i)
 \Equals*[distributing parallel over non-deterministic choice (\refaxiom*{parallel-distribute})]
  ((c \quotient i) \nondet (d \quotient i)) ~\parallel~ i 
\end{refine}%
\end{proof}
Distribution of the rely quotient over parallel requires
a proviso on the interference $i$ that $i \parallel i ~\refsto~i$.
That distribution law follows from a more general law with a parallel in both arguments of the quotient.
\begin{lawx}[rely-distribute-parallel]
\begin{eqnarray}
  (c \parallel d) \quotient (i \parallel j)  & ~\refsto~ & (c \quotient i) ~\parallel~ (d \quotient j) \labelproperty{rely-distribute-parallel-a} \\
  (c \parallel d) \quotient i                    & ~\refsto~ & (c \quotient i) ~\parallel~ (d \quotient i)  \mbox{~~~if $i \parallel i ~\refsto~ i$}
     \labelproperty{rely-distribute-parallel-b}
\end{eqnarray}
\end{lawx}

\begin{proof}
By \Law{rely-refinement}, (\refproperty*{rely-distribute-parallel-a}) holds if 
$c \parallel d ~\refsto~ (c \quotient i) \parallel (d \quotient j) \parallel i \parallel j$,
which holds as follows.
\begin{refine}
  c \parallel d
 \Refsto*[by \Law{rely-quotient} twice]
  ((c \quotient i) \parallel i) \parallel ((d \quotient j) \parallel j)
 \Equals*[by associativity (\refaxiom*{parallel-associative}) and commutativity (\refaxiom*{parallel-commutes}) of parallel]
  (c \quotient i) \parallel (d \quotient j) \parallel i \parallel j
\end{refine}%
The proof of (\refproperty*{rely-distribute-parallel-b}) uses (\refproperty*{rely-distribute-parallel-a}) with $j = i$ as follows.
\begin{refine}
  (c \parallel d) \quotient i
 \Refsto*[by \Law{rely-weaken} using assumption $i \parallel i \refsto i$] 
  (c \parallel d) \quotient (i \parallel i)
 \Refsto*[by part (\refproperty*{rely-distribute-parallel-a}) with $j = i$]
  (c \quotient i) ~\parallel~ (d \quotient i)
\end{refine}%
\end{proof}
\begin{relational}%
For a relational rely condition,
if $i = \FinGuar{r}$ then by (\refproperty*{parallel-atomic-kleene}),
\(
  \FinGuar{r} \parallel \FinGuar{r}  ~=~  \FinGuar{r \cup r} ~=~  \FinGuar{r}, 
\)
and hence the proviso for (\refproperty*{rely-distribute-parallel-b}) holds in this case.
The fact that the proviso for a relational rely condition holds
allows rely conditions to be distributed into any parallel composition.\\
\end{relational}
Distribution of a rely quotient of a process $i$ over a sequential composition 
requires that separate occurrences of $i$ running in parallel with each command in the sequence
can be refined to a single occurrence of $i$ run in parallel with the sequence
as given by condition (\refproperty*{rely-distribute-sequential-assumption}).
\begin{lawx}[rely-distribute-sequential]
If for process $i$, 
\begin{eqnarray}\labelproperty{rely-distribute-sequential-assumption}
  \forall c_0, c_1 \spot 
    (c_0 \parallel i) \SSeq (c_1 \parallel i) ~\refsto~ (c_0 \SSeq c_1) \parallel i,
\end{eqnarray}
then~~~
\begin{eqnarray}\labelproperty{rely-distribute-sequential}
  (c \SSeq d) \quotient i  & ~\refsto~ & (c \quotient i) \SSeq (d \quotient i)~.
\end{eqnarray}
\end{lawx}

\begin{proof}
By \Law{rely-refinement}, (\refproperty*{rely-distribute-sequential}) is equivalent to
$c \SSeq d \refsto ((c \quotient i) \SSeq (d \quotient i)) \parallel i$.
\begin{refine}
    c \SSeq d
  \Refsto*[by \Law{rely-quotient} twice]
    ((c \quotient i) \parallel i) \SSeq ((d \quotient i) \parallel i)
  \Refsto*[by assumption (\refproperty*{rely-distribute-sequential-assumption}) with $c_0 = c \quotient i$ and $c_1 = d \quotient i$]
    ((c \quotient i) \SSeq (d \quotient i)) \parallel i
\end{refine}
\end{proof}

\begin{relational}%
For a relational rely condition,
if $i = \FinGuar{r}$ then 
\(
  (c \parallel \FinGuar{r}) \SSeq (d \parallel \FinGuar{r})  =  (c \SSeq d) \parallel \FinGuar{r}
\)
holds for any $c$, $d$ and $r$ and hence proviso (\refproperty*{rely-distribute-sequential-assumption}) holds.
As with parallel, 
the use of a relational rely condition allows the rely to be distributed into any sequential composition.
In the general case, if proviso (\refproperty*{rely-distribute-sequential-assumption}) does not hold
the question arises as to what alternative approaches could be used -- as with \Law{rely-distribute-parallel}
these are likely to depend on the form of the interference.\\
\end{relational}

Distribution of the rely quotient over an iteration requires
the same side condition (\refproperty*{rely-distribute-sequential-assumption}) on distribution of the interference $i$ over a sequential composition
as for \Law*{rely-distribute-sequential}.
The law uses the more general form $c\FinOrInfIter \SSeq d = \mu x \spot d \nondet c \SSeq x$.
This allows the law to be applied to a while loop $\While b \Do c$,
which can be defined in the form $(bc)\FinOrInfIter\bar{b}$
where $b$ stands for the test of the while loop succeeding and $\bar{b}$ for it failing.
Just developing a law for $c\FinOrInfIter$ is problematic for the zero iterations case 
because this corresponds to $\Nil \quotient i$ and $\Nil \quotient i \refsto d$ holds if and only if
$\Nil \refsto d \parallel i$, which only holds if $i$ behaves as either $\Nil$ or $\Magic$.
\begin{lawx}[rely-distribute-iteration]
If 
\begin{eqnarray}\labelproperty{rely-distribute-iteration-assumption}
  \forall c_0, c_1 \spot 
    (c_0 \parallel i) \SSeq (c_1 \parallel i) ~\refsto~ (c_0 \SSeq c_1) \parallel i,
\end{eqnarray}
holds for $i$,~~~
\(
  (c\FinOrInfIter \SSeq d) \quotient i ~~\refsto~~ (c \quotient i)\FinOrInfIter \SSeq (d \quotient i)~.
\)
\end{lawx}

\begin{proof}
By \Law{rely-refinement} the law is equivalent to 
$c\FinOrInfIter \SSeq d ~\refsto~ ((c \quotient i)\FinOrInfIter \SSeq (d \quotient i)) \parallel i $
and by \Lemma{induction} it is sufficient to show,
\begin{eqnarray*}
  d \nondet c \SSeq (((c \quotient i)\FinOrInfIter \SSeq (d \quotient i)) \parallel i) ~\refsto~ ((c \quotient i)\FinOrInfIter \SSeq (d \quotient i)) \parallel i,
\end{eqnarray*}
which can be shown as follows.
\begin{refine}
  d \nondet c \SSeq (((c \quotient i)\FinOrInfIter \SSeq (d \quotient i)) \parallel i)
 \Refsto*[by \Law{rely-quotient} applied to each of the first $d$ and $c$]
  ((d \quotient i) \parallel i) \nondet ((c \quotient i) \parallel i) \SSeq (((c \quotient i)\FinOrInfIter \SSeq (d \quotient i)) \parallel i)
 \Refsto*[by assumption (\refproperty*{rely-distribute-iteration-assumption}) with $c_0 = c \quotient i$ and $c_1 = (c \quotient i)\FinOrInfIter \SSeq (d \quotient i)$]
  ((d \quotient i) \parallel i) \nondet (((c \quotient i) \SSeq (c \quotient i)\FinOrInfIter \SSeq (d \quotient i)) \parallel i)
 \Equals*[distributing parallel over non-deterministic choice (\refaxiom*{parallel-distribute})]
  ((d \quotient i) \nondet (c \quotient i) \SSeq (c \quotient i)\FinOrInfIter \SSeq (d \quotient i)) \parallel i
 \Equals*[factoring out $d \quotient i$ using (\refaxiom*{sequential-distribute-nondet-right})]
  ((\Nil \nondet (c \quotient i) \SSeq (c \quotient i)\FinOrInfIter) \SSeq (d \quotient i)) \parallel i
 \Equals*[folding using (\refaxiom*{least-fixed-point-unfold})]
  ((c \quotient i)\FinOrInfIter \SSeq (d \quotient i)) \parallel i
\end{refine}%
\end{proof}

\begin{relational}%
The proviso (\refproperty*{rely-distribute-iteration-assumption}) holds for a relational rely $i = \FinGuar{r}$
and hence \Law{rely-distribute-iteration} holds in this case.
\end{relational}

The following laws combine distribution properties with the introduction of a parallel composition.
\begin{lawx}[parallel-introduce-with-rely]~~
\(
  (c \together d) \quotient i ~~\refsto~ \begin{array}[t]{l}
                                                              (j_1 \together (c \quotient (j_0 \parallel i))) ~\parallel~ 
                                                              (j_0 \together (d \quotient (j_1 \parallel i)))
                                                             \end{array}
\)
\end{lawx}

\begin{proof}\mbox{}
\begin{refine}
  (c \together d) \quotient i
 \Refsto*[by \Law{rely-distribute-conjunction}]
  (c \quotient i) ~\together~ (d \quotient i)
 \Refsto*[by \Law{parallel-introduce}]
  (j_1 \together ((c \quotient i) \quotient j_0)) ~\parallel~ (j_0 \together ((d \quotient i) \quotient j_1))
 \Equals*[by \Law{rely-nested} twice]
  (j_1 \together (c \quotient (j_0 \parallel i))) ~\parallel~ (j_0 \together (d \quotient (j_1 \parallel i)))
\end{refine}%
\end{proof}
In the right side of the above law one branch of the parallel guarantees $j_1$ and the other guarantees $j_0$,
and hence their parallel combination guarantees $j_1 \parallel j_0$.

\begin{lawx}[parallel-introduce-with-rely-guarantee]
\begin{eqnarray*}
  (j_1 \parallel j_0) ~\together~ (c \together d) \quotient i & ~~\refsto~ & \begin{array}[t]{l}
                                                              (j_1 \together (c \quotient (j_0 \parallel i))) ~\parallel~ 
                                                              (j_0 \together (d \quotient (j_1 \parallel i)))~.
                                                             \end{array}
\end{eqnarray*}
\end{lawx}

\begin{proof}\mbox{}
\begin{refine}
  (j_1 \parallel j_0) ~\together~ ((c \together d) \quotient i)
 \Refsto*[by \Law{parallel-introduce-with-rely}]
  (j_1 \parallel j_0) ~\together~ ((j_1 \together (c \quotient (j_0 \parallel i))) ~\parallel~ (j_0 \together (d \quotient (j_1 \parallel i))))
 \Refsto*[exchanging \strictconjunction\ and parallel by \refaxiom{conjunction-exchange-parallel}]
  (j_1 \together j_1 \together (c \quotient (j_0 \parallel i))) ~\parallel~ (j_0 \together j_0 \together (d \quotient (j_1 \parallel i)))
 \Equals*[as \strictconjunction\ is idempotent (\refaxiom*{conjunction-idempotent})]
  (j_1 \together (c \quotient (j_0 \parallel i))) ~\parallel~ (j_0 \together (d \quotient (j_1 \parallel i)))
\end{refine}%
\end{proof}

\begin{relational}%
In the relational model by (\refproperty*{parallel-atomic-omega2}),
$\InfGuar{g \cup r} \refsto \InfGuar{g} \parallel \InfGuar{r}$
and hence if $j_1 = \InfGuar{g}$ and $j_0 = \InfGuar{r}$ the effective guarantee for \Law*{parallel-introduce-with-rely-guarantee}
is $g \cup r$.\\
\vspace*{-2ex}%
\end{relational}

\section{Relationship to relational rely}\label{section:Jones-rely}

This section explores the relationship to the Jones-style rely condition.
Jones considered total correctness rules for handling the implementation of a
pre-post specification in a context satisfying a rely condition \cite{CoJo07}.
To instantiate the general theory presented here for Jones-style rely-guarantee rules,
termination needs to be handled.
For a terminating command, such as a specification $\Spec{}{}{q}$, using a rely quotient of
$\Spec{}{}{q} \quotient \InfGuar{r}$ leads to an infeasible specification 
because by \Law{rely-quotient} this requires
\begin{eqnarray*}
  \Spec{}{}{q} ~\refsto~ (\Spec{}{}{q} \quotient \InfGuar{r}) ~\parallel~ \InfGuar{r}
\end{eqnarray*}
but $\Spec{}{}{q}$ is terminating and $\InfGuar{r}$ has non-terminating behaviours
and hence $\Spec{}{}{q} \quotient \InfGuar{r}$ must rule out such infinite behaviours of its environment.
However, executable code cannot rule out behaviours of its environment and
hence using $\InfGuar{r}$ for a rely quotient for a terminating command 
is not a feasible approach.
Therefore the terminating iteration $\FinGuar{r}$ must be used.
Choosing $i$ and $j$ be the processes $\FinGuar{r}$ and $\FinGuar{g}$, respectively,
in \Law{parallel-introduce} gives the following.
\begin{eqnarray}
  \labelproperty{parallel-introduce-relational}
  c \together d & ~~\refsto & \begin{array}[t]{l}
                                          (\FinGuar{g} \together (c \quotient \FinGuar{r})) ~\parallel~
                                          (\FinGuar{r} \together (d \quotient \FinGuar{g}))
                                         \end{array}
\end{eqnarray}
Note that due to the use of a \strictconjunction\ to enforce a guarantee,
the first branch of the parallel composition is only required to maintain its guarantee condition $g$
as long as its environment maintains its rely condition $r$.
If its environment does not maintain $r$ the rely quotient can abort,
at which point the whole branch of the parallel is considered to have aborted
and hence the guarantee no longer needs to be maintained.

The parallel introduction rule of Jones \cite{jon83a} takes a postcondition of the form
$q_0 \relint q_1$ and introduces a parallel composition in which the two branches
ensure $q_0$ and $q_1$ respectively.
\begin{lawx}[parallel-specification]
\begin{eqnarray*}
    \Pre{p} \SSeq \Spec{}{}{q_0 \relint q_1} & ~\refsto & 
      \begin{array}[t]{l}
        (\Pre{p} \SSeq (\FinGuar{g} \together (\Spec{}{}{q_0} \quotient \FinGuar{r}))) \parallel 
        (\Pre{p} \SSeq (\FinGuar{r} \together (\Spec{}{}{q_1} \quotient \FinGuar{g})))
      \end{array}
\end{eqnarray*}
\end{lawx}

\begin{proof}
Note that by (\refproperty*{spec-conjoin-spec}) 
a specification $\Spec{}{}{q_0 \relint q_1}$ is equivalent to $\Spec{}{}{q_0} \together \Spec{}{}{q_1}$.
\begin{refine}
    \Pre{p} \SSeq \Spec{}{}{q_0 \relint q_1} 
  \Equals*[by (\refproperty*{spec-conjoin-spec})]
    \Pre{p} \SSeq (\Spec{}{}{q_0} \together \Spec{}{}{q_1}) 
  \Refsto*[by \Law{parallel-introduce}]
    \Pre{p} \SSeq
    ((\FinGuar{g} \together (\Spec{}{}{q_0} \quotient \FinGuar{r})) ~\parallel~ 
     (\FinGuar{r} \together (\Spec{}{}{q_1} \quotient \FinGuar{g})))
  \Equals*[by \Law{precondition-parallel}]
    (\Pre{p} \SSeq (\FinGuar{g} \together (\Spec{}{}{q_0} \quotient \FinGuar{r}))) ~\parallel~ 
    (\Pre{p} \SSeq (\FinGuar{r} \together (\Spec{}{}{q_1} \quotient \FinGuar{g})))
\end{refine}
\end{proof}
The above corresponds to the Jones-style proof rule for introducing a parallel composition
although phrased in refinement calculus form rather than as a quintuple.

\section{Fair parallelism}\label{section:fair-parallel}

This section highlights the parts of the theory that are influenced by
the choice as to whether or not parallelism is assumed to be fair.
The semantics for parallel does not require fairness.
A fair semantics would rule out traces ending in an infinite sequence of
program steps of one process, if the other process could make a program step.
Most algebraic properties are independent of whether or not parallel 
is assumed to be fair.
Fair parallel is denoted by $c \parallel_f d$.
It refines the parallel operator used so far, which does not assume fairness.
\begin{eqnarray}
  c \parallel d & ~\refsto~ & c \parallel_f d
    \labelproperty{fair-refines-parallel}
\end{eqnarray}
If no fairness assumption is made about the parallel operator,
the notion of termination of a process is weak as it means 
a process terminates provided it is not permanently interrupted by its environment.
For the program
\begin{refine}
  x := 1;
  ((\While x \neq 0 \Do \Skip) \parallel x := 0)
\end{refine}%
the while loop will not terminate unless the $x := 0$
is given a chance to set $x$ to 0.
If parallelism is not assumed to be fair, 
the loop is not guaranteed to terminate even if it is not permanently interrupted;
in fact the problem comes if it is never interrupted by $x := 0$.
However, if parallel is assumed to be fair,
the right process will eventually set $x$ to 0 and the loop will terminate.

Because the definition of the rely quotient operator depends on the parallel operator
there is different quotient operator corresponding to fair parallel.
\begin{definitionx}[fair-quotient]~~~
\(
  c \quotient_f i ~~\sdefs~~ \Nondet \{ d \where (c \refsto d \parallel_f i) \}
\)
\end{definitionx}
From (\refproperty*{fair-refines-parallel}) it follows that 
$c \quotient_f i ~\refsto~ c \quotient i$,
that is, any implementation that handles any interference from process $i$
also handles fair interference from process $i$.

In the relational model, the property
\begin{eqnarray}
  \InfGuar{r_0 \cup r_1} & ~=~ & \InfGuar{r_0} \parallel \InfGuar{r_1}
    \labelproperty{parallel-atomic-omega3}
\end{eqnarray}
holds, but if parallel is fair (\refproperty*{parallel-atomic-omega3}) becomes a refinement 
because the left command allows an infinite sequence of steps satisfying $r_0$
(that do not satisfy $r_0 \cap r_1$),
while the right command does not allow such a sequence if parallel is fair.
In proving the laws in this paper, we have relied on (\refproperty*{parallel-atomic-omega3}) only being a refinement, 
i.e.\ property (\refproperty*{parallel-atomic-omega2}),
and hence our laws also apply for fair parallel and fair quotient.

\section{Related work}

Dingel developed a refinement calculus for rely-guarantee concurrency \cite{DingelPhD,Dingel02}.
Like \cite{HayesJonesColvin14TR} it is based on relational rely and guarantee conditions
but unlike \cite{HayesJonesColvin14TR} and here, 
it makes use of a monolithic specification which is a four-tuple of pre, rely, guarantee and post conditions,
rather than our separate commands and operators.
The approach used here has the benefit of separating the different concepts 
and providing laws for each operator as well as combinations of operators.
The laws given here can be combined to derive laws similar to those of Dingel as well as many other laws.
The other major advance over Dingel is the generalisation to use processes 
for relies and guarantees.

Hoare {\it et al.}\ \cite{DBLP:journals/jlp/HoareMSW11} have developed a \emph{Concurrent Kleene Algebra (CKA)} 
and investigated its extension to a \emph{rely/guarantee CKA}.
Their algebra includes the axiom $(c \SSeq \Magic) = \Magic$,
which is not satisfied if $c$ is either a non-terminating process or $\Abort$
and hence they only consider partial correctness.
The rely/guarantee CKA includes a sub-algebra of commands called \emph{invariants},
in which an invariant $j$ satisfies
\begin{eqnarray}
  j & \refsto & \Nil \labelproperty{invariant-refsto-nil} \\
  j & \refsto & j \parallel j \labelproperty{invariant-refsto-parallel} \\
  j & \refsto & j \SSeq j \labelproperty{invariant-refsto-sequential}
\end{eqnarray}
because in their algebra $c \parallel d ~\refsto~ c \SSeq d$ and hence 
(\refproperty*{invariant-refsto-parallel}) implies (\refproperty*{invariant-refsto-sequential}).
Properties (\refproperty*{invariant-refsto-parallel}) and (\refproperty*{invariant-refsto-sequential}) match
the properties 
used in \Law{conjunction-distribute} parts (\refproperty*{conjunction-distribute-parallel}) and (\refproperty*{conjunction-distribute-sequential}).
Properties (\refproperty*{invariant-refsto-nil}) and (\refproperty*{invariant-refsto-sequential}) together 
ensure that $j = j\FinIter$ and hence that 
$j \together d\FinIter = j\FinIter \together d\FinIter \refsto (j \together d)\FinIter$
matching \Law{conjunction-distribute} part (\refproperty*{conjunction-distribute-kleene}).
In a rely/guarantee CKA, for any $c$ and $d$ and any invariant $j$, 
\begin{eqnarray*}
  (c \parallel j) \SSeq (d \parallel j) & ~\refsto~ & (c \SSeq d) \parallel j~,
\end{eqnarray*}
which matches our property (\refproperty*{rely-distribute-sequential-assumption}).
A rely/guarantee CKA does not require our property $j \parallel j \refsto j$ 
but \cite{DBLP:journals/jlp/HoareMSW11} does not consider an equivalent of \Law{rely-distribute-parallel}
for which this property is required.
In a rely/guarantee CKA 
a Jones-like rely-guarantee quintuple,
written $p\,r \{ d \} c\,g$ there, 
is defined in terms of a Hoare triple plus guarantee condition,
in which $r$ and $g$ are invariants (rather than relations).
\begin{eqnarray}
  p\,r \{ d\} c\,g & ~\sdefs~ &  p \{ r \parallel d \} c ~\land~d \mathrel{\Keyword{guar}} g~,
    \labelproperty{CKA-rg}
\end{eqnarray}
Our ``equivalent'' of (\refproperty*{CKA-rg}) is of the form
\begin{eqnarray}
  g \together \Pre{p} \SSeq (c \quotient r) & ~\refsto~ & d~,
    \labelproperty{gen-rg}
\end{eqnarray}
although the two differ due to the different approaches taken.
Because $g$ is an invariant 
the requirement $d \mathrel{\Keyword{guar}} g$ in (\refproperty*{CKA-rg}) reduces to $g \refsto d$,
which is stronger than the requirement in (\refproperty*{gen-rg}). 
Firstly, in (\refproperty*{gen-rg}) $d$ is only required to satisfy the guarantee from initial states satisfying the precondition $p$.
Secondly and more subtly, $c \quotient r$ may abort 
because its environment does not satisfy $r$ and 
hence the left side of (\refproperty*{gen-rg}) aborts and so $d$ no longer needs to maintain the guarantee.
This latter condition corresponds to Jones' requirement that the implementation only needs to maintain
the guarantee condition as long as its environment maintains the rely condition~\cite{jon83a}.
Our ability to use the weaker requirement comes from the use of the \strictconjunction\ operator,
which is not available in CKA.

\section{Conclusions}\label{section:conclusions}

The main contribution of this paper is to explore the essence of the rely-guarantee approach to concurrency.
Jones' guarantee condition is generalised from a relation to a process 
by making use of a \strictconjunction\ operator
and his rely condition from a relation to a process by introducing a rely quotient operator,
which forms a residual with respect to parallel composition (see \Law{rely-refinement}).
Both \strictconjunction\ and rely quotient have simple algebraic properties.
The \strictconjunction\ operator and parallel composition satisfy an exchange property
(\refaxiom*{conjunction-exchange-parallel})
which leads to a simple and elegant proof of \Law{parallel-introduce}, 
which is the key law for introducing a parallel composition in the generalised rely-guarantee theory.
Because our theory allows non-terminating processes,
it can handle total correctness properties as well as reasoning about non-terminating processes.

Generalising rely-guarantee theory so that guarantees and relies are arbitrary processes
rather than binary relations has highlighted the important algebraic properties of rely-guarantee theory.
In \Law{conjunction-distribute}, for a \strictconjunction\ of a command 
to distribute over a parallel composition one needs proviso (\refproperty*{c-to-c-parallel-c}); 
to distribute over a sequential composition one needs (\refproperty*{c-to-c-seq-c}); 
and
to distribute over finite 
iteration one needs 
(\refproperty*{c-to-c-seq-c}) and (\refproperty*{c-nil}).
\begin{eqnarray}
  c & ~\refsto~ & c\parallel c
    \labelproperty{c-to-c-parallel-c} \\
  c & ~\refsto~ & c \SSeq c
    \labelproperty{c-to-c-seq-c} \\
  c & ~\refsto~ & \Nil
    \labelproperty{c-nil} 
\end{eqnarray}
Because all these properties hold if $c$ is of the form $\InfGuar{g}$ for any relation $g$,
the choice by Jones to represent interference by an (iterated atomic) relation,
rather than a general process,
means that \Law{guarantee-distribute} for the relational model 
does not require any provisos.

Even within the relational model more expressive guarantees are possible,
for example, a guarantee of $\FinGuar{g_0} \SSeq \FinGuar{g_1}$ on $c$
may lead to the following refinement, in which $c$ is refined sequentially to match the guarantees.
\begin{refine}
  \FinGuar{g_0} \SSeq \FinGuar{g_1} \together c
 \Refsto*[assuming $c \refsto c_0 \SSeq c_1$]
  \FinGuar{g_0} \SSeq \FinGuar{g_1} \together c_0 \SSeq c_1
 \Refsto*[exchanging \strictconjunction\ and sequential composition (\refaxiom*{conjunction-exchange-sequential})]
  (\FinGuar{g_0} \together c_0) \SSeq (\FinGuar{g_1} \together c_1)
\end{refine}

\Law{rely-distribute-parallel} has a proviso of (\refproperty*{i-parallel-i-to-i}),
and both \Law{rely-distribute-sequential} and \Law{rely-distribute-iteration} have a proviso of 
(\refproperty*{c-par-i-seq-d-par-i-to-c-seq-d-par-i}).
\begin{eqnarray}
  i \parallel i & ~\refsto~ & i 
    \labelproperty{i-parallel-i-to-i} \\ 
  \forall c_0, c_1 \spot (c_0 \parallel i) \SSeq (c_1 \parallel i)  & ~\refsto~ &  (c_0 \SSeq c_1) \parallel i
    \labelproperty{c-par-i-seq-d-par-i-to-c-seq-d-par-i}
\end{eqnarray}
Because both these properties hold for $i$ of the form $\FinGuar{r}$ for any relation $r$,
the laws do not require any provisos for relational rely conditions
thus simplifying the process of distributing relational rely conditions.
Note that taking $c_0$ and $c_1$ to both be $\Skip$ in (\refproperty*{c-par-i-seq-d-par-i-to-c-seq-d-par-i})
gives $i \SSeq i \refsto i$.
An interesting question for future research is what other processes satisfy the
provisos required for the distribution properties to hold, 
or what other distribution properties can be used in their place.

In this paper we have considered an example model based on relational rely-guarantee. 
The model is similar to that used by others \cite{CoJo07,BoerHannemanDeRoever99,Dingel02,DeRoever01,HayesJonesColvin14TR}
but even within the relational model, guarantees and relies are treated more generally as processes.
Other possible models for future consideration are an event-based model
similar to that used with Concurrent Kleene Algebra \cite{DBLP:journals/jlp/HoareMSW11}
or a model that handles concurrency in a hybrid setting.

\section*{Acknowledgements}

The research reported here was supported by
Australian Research Council Grant DP130102901.
This paper has benefited from feedback from
Robert Colvin,
Cliff Jones,
Jo\~{a}o Ferreira,
Larissa Meinicke, 
Carroll Morgan,
Kim Solin,
Georg Struth,
Kirsten Winter
and the anonymous referees 
but the remaining errors are all courtesy of the author.
Special thanks go to Julian Fell and Andrius Velykis for mechanising the proofs of the laws in Isabelle/HOL.

\bibliographystyle{alpha}
\bibliography{absrg}

\newcommand{\etalchar}[1]{$^{#1}$}
\begin{thebibliography}{dBHdR99}

\bibitem[Aar92]{Aarts92}
C.~J. Aarts.
\newblock Galois connections presented calculationally.
\newblock Technical report, Department of Computing Science, Eindhoven
  University of Technology, 1992.
\newblock Afstudeer verslag (Graduating Dissertation).

\bibitem[ABB{\etalchar{+}}95]{fixedpointcalculus1995}
Chritiene Aarts, Roland Backhouse, Eerke Boiten, Henk Doombos, Netty van
  Gasteren, Rik van Geldrop, Paul Hoogendijk, Ed~Voermans, and Jaap van~der
  Woude.
\newblock Fixed-point calculus.
\newblock {\em Information Processing Letters}, 53:131--136, 1995.
\newblock {Mathematics of Program Construction Group.}

\bibitem[Acz83]{Aczel83}
P.~H.~G. Aczel.
\newblock On an inference rule for parallel composition, 1983.
\newblock Private communication to Cliff Jones
  \url{http://homepages.cs.ncl.ac.uk/cliff.jones/publications/MSs/PHGA-traces.pdf}.

\bibitem[Bac81]{Bac81a}
R.-J.~R. Back.
\newblock On correct refinement of programs.
\newblock {\em Journal of Computer and System Sciences}, 23(1):49--68, February
  1981.

\bibitem[BCG02]{Backhouse02}
Roland Backhouse, Roy Crole, and Jeremy Gibbons, editors.
\newblock {\em Algebraic and Coalgebraic Methods in the Mathematics of Program
  Construction}.
\newblock Springer, 2002.

\bibitem[Bli78]{Blikle78}
Andrzej Blikle.
\newblock Specified programming.
\newblock In Edward~K. Blum, Manfred Paul, and Satoru Takasu, editors, {\em
  Mathematical Studies of Information Processing}, volume~75 of {\em Lecture
  Notes in Computer Science}, pages 228--251. Springer, 1978.

\bibitem[BvW98]{BackWright98}
R.-J.~R. Back and J.~von Wright.
\newblock {\em Refinement Calculus: A Systematic Introduction}.
\newblock Springer, New York, 1998.

\bibitem[BvW99]{BackWright99}
R.-J.R. Back and J.~von Wright.
\newblock Reasoning algebraically about loops.
\newblock {\em Acta Informatica}, 36:295--334, 1999.

\bibitem[CH81]{ChaochenHoare81}
Zhou Chaochen and C.~A.~R. Hoare.
\newblock Partial correctness of communication protocols.
\newblock In {\em Technical Monograph PRG-20, Partial Correctness of
  Communicating Processes and Protocols}, pages 13--23. Oxford University
  Computing Laboratory, May 1981.

\bibitem[Cha82]{Chaochen:1982}
Zhou Chaochen.
\newblock Weakest environment of communicating processes.
\newblock In {\em Proc.\ of the June 7-10, 1982, National Computer Conf.},
  AFIPS '82, pages 679--690, New York, NY, USA, 1982. ACM.

\bibitem[CJ07]{CoJo07}
J.~W. Coleman and C.~B. Jones.
\newblock A structural proof of the soundness of rely/guarantee rules.
\newblock {\em Journal of Logic and Computation}, 17(4):807--841, 2007.

\bibitem[Con71]{conway71}
J.H. Conway.
\newblock {\em Regular Algebra and Finite Machines}.
\newblock Chapman \& Hall, 1971.

\bibitem[dBHdR99]{BoerHannemanDeRoever99}
F.S. de~Boer, U.~Hannemann, and W.-P. de~Roever.
\newblock Formal justification of the rely-guarantee paradigm for
  shared-variable concurrency: a semantic approach.
\newblock In Jeannette Wing, Jim Woodcock, and Jim Davies, editors, {\em
  FM’99 — Formal Methods}, volume 1709 of {\em Lecture Notes in Computer
  Science}, pages 714--714. Springer Berlin / Heidelberg, 1999.

\bibitem[Din00]{DingelPhD}
J{\"u}rgen Dingel.
\newblock {\em Systematic Parallel Programming}.
\newblock PhD thesis, Carnegie Mellon University, 2000.
\newblock CMU-CS-99-172.

\bibitem[Din02]{Dingel02}
J.~Dingel.
\newblock A refinement calculus for shared-variable parallel and distributed
  programming.
\newblock {\em Formal Aspects of Computing}, 14(2):123--197, 2002.

\bibitem[dR01]{DeRoever01}
W.-P. de~Roever.
\newblock {\em Concurrency Verification: Introduction to Compositional and
  Noncompositional Methods}.
\newblock Cambridge University Press, 2001.

\bibitem[HH86]{HoareHe86}
C.A.R. Hoare and Jifeng He.
\newblock The weakest prespecification.
\newblock {\em Fundamenta Informaticae}, IX:51--84, 1986.

\bibitem[HHH{\etalchar{+}}87]{HoareHayesEtcFull87}
C.~A.~R. Hoare, I.~J. Hayes, {He Jifeng}, C.~Morgan, A.~W. Roscoe, J.~W.
  Sanders, I.~H. S{\o}rensen, J.~M. Spivey, and B.~A. Sufrin.
\newblock Laws of programming.
\newblock {\em Communications of the ACM}, 30(8):672--686, August 1987.
\newblock Corrigenda: CACM 30(9):770.

\bibitem[HJC14]{HayesJonesColvin14TR}
Ian~J. Hayes, Cliff~B. Jones, and Robert~J. Colvin.
\newblock Laws and semantics for rely-guarantee refinement.
\newblock Technical Report CS-TR-1425, Newcastle University, July 2014.

\bibitem[HMSW11]{DBLP:journals/jlp/HoareMSW11}
Tony Hoare, B.~M{\"o}ller, G.~Struth, and I.~Wehrman.
\newblock {Concurrent Kleene Algebra} and its foundations.
\newblock {\em J. Log. Algebr. Program.}, 80(6):266--296, 2011.

\bibitem[Hoa69]{Hoare69a}
C.~A.~R. Hoare.
\newblock An axiomatic basis for computer programming.
\newblock {\em Communications of the ACM}, 12(10):576--580, 583, October 1969.

\bibitem[JHC15]{FACJexSEFM-14}
Cliff~B. Jones, Ian~J. Hayes, and Robert~J. Colvin.
\newblock Balancing expressiveness in formal approaches to concurrency.
\newblock {\em Formal Aspects of Computing}, 27(3):475--497, May 2015.

\bibitem[Jon81]{Jones81d}
C.~B. Jones.
\newblock {\em Development Methods for Computer Programs including a Notion of
  Interference}.
\newblock PhD thesis, Oxford University, June 1981.
\newblock Printed as: Programming Research Group, Technical Monograph 25.

\bibitem[Jon83]{jon83a}
C.B. Jones.
\newblock Tentative steps toward a development method for interfering programs.
\newblock {\em ACM Transactions on Programming Languages and Systems},
  5(4):596--619, October 1983.

\bibitem[Jon96]{jones96a}
C.~B. Jones.
\newblock Accommodating interference in the formal design of concurrent
  object-based programs.
\newblock {\em Formal Methods in System Design}, 8(2):105--122, March 1996.

\bibitem[Koz97]{kozen97kleene}
Dexter Kozen.
\newblock Kleene algebra with tests.
\newblock {\em ACM Trans.\ Prog.\ Lang.\ and Sys.}, 19(3):427--443, May 1997.

\bibitem[Mor87]{ATBfSRatPC}
J.~M. Morris.
\newblock A theoretical basis for stepwise refinement and the programming
  calculus.
\newblock {\em Science of Computer Programming}, 9(3):287--306, 1987.

\bibitem[Mor88]{TSS}
C.~C. Morgan.
\newblock The specification statement.
\newblock {\em ACM Trans.\ Prog.\ Lang.\ and Sys.}, 10(3):403--419, July 1988.

\bibitem[Mor94]{Morgan94}
C.~C. Morgan.
\newblock {\em Programming from Specifications}.
\newblock Prentice Hall, second edition, 1994.

\bibitem[vW04]{Wright04}
J.~von Wright.
\newblock Towards a refinement algebra.
\newblock {\em Sci. of Comp. Prog.}, 51:23--45, 2004.

\end{thebibliography}

\newpage
{\small
\printindex
}

\end{document}